\documentclass{article} 
\usepackage{iclr2024_conference,times}
\usepackage{array}
\usepackage{hyperref}
\usepackage{url}
\usepackage{microtype}
\usepackage{graphicx}
\usepackage{subfigure}
\usepackage{amssymb}
\usepackage{amsfonts}
\usepackage{fdsymbol}
\usepackage{booktabs} 
\usepackage{multirow}
\usepackage{amsmath}
\usepackage{threeparttable}
\usepackage{appendix}
\usepackage{ulem}
\usepackage{wrapfig}
\usepackage{caption}
\usepackage{adjustbox}
\usepackage{subfigure}
\usepackage{tikz}
\usepackage{fontawesome}
\usepackage{threeparttable}
\usepackage{algorithmic}
\usepackage{algorithm}
\usepackage{arydshln}
\usepackage{booktabs}
\definecolor{orange}{HTML}{FF7F00}

\DeclareMathOperator*{\argmin}{arg\,min}

\usepackage{amsmath,amsfonts,bm}

\def\eqref#1{equation~\ref{#1}}

\def\1{\bm{1}}

\DeclareMathAlphabet{\mathsfit}{\encodingdefault}{\sfdefault}{m}{sl}
\SetMathAlphabet{\mathsfit}{bold}{\encodingdefault}{\sfdefault}{bx}{n}

\usepackage{hyperref}
\usepackage{url}

\title{BELT-2: Bootstrapping EEG-to-Language representation alignment for multi-task brain decoding}

\iclrfinalcopy
\author{Jinzhao Zhou, Yiqun Duan, Fred Chang, Thomas Do, Yu-Kai Wang,  Chin-Teng Lin}
\begin{document}

\maketitle

\begin{abstract}
The remarkable success of large language models (LLMs) across various multi-modality applications is well established. However, integrating large language models with humans, or brain dynamics, remains relatively unexplored. In this paper, we introduce BELT-2, a pioneering multi-task model designed to enhance both encoding and decoding performance from EEG signals. To bolster the quality of the EEG encoder, BELT-2 is the first work to innovatively 1) adopt byte-pair encoding (BPE)-level EEG-language alignment and 2) integrate multi-task training and decoding in the EEG domain. Inspired by the idea of \textbf{\textit{Bridging the Brain with GPT}}, we further connect the multi-task EEG encoder with LLMs by utilizing prefix-tuning on intermediary output from the EEG encoder. These innovative efforts make BELT-2 a pioneering breakthrough, making it the first work in the field capable of decoding coherent and readable sentences from non-invasive brain signals. Our experiments highlight significant advancements over prior techniques in both quantitative and qualitative measures, achieving a decoding performance with a BLEU-1 score of 52.2\% on the ZuCo dataset. Furthermore, BELT-2 shows a remarkable improvement ranging from 31\% to 162\% on other translation benchmarks. Codes can be accessed via the provided anonymous link~\footnote{https://anonymous.4open.science/r/BELT-2-0048}.
\end{abstract}

\begin{figure}[hbpt]
    \centering
    \includegraphics[width=1.0\textwidth]{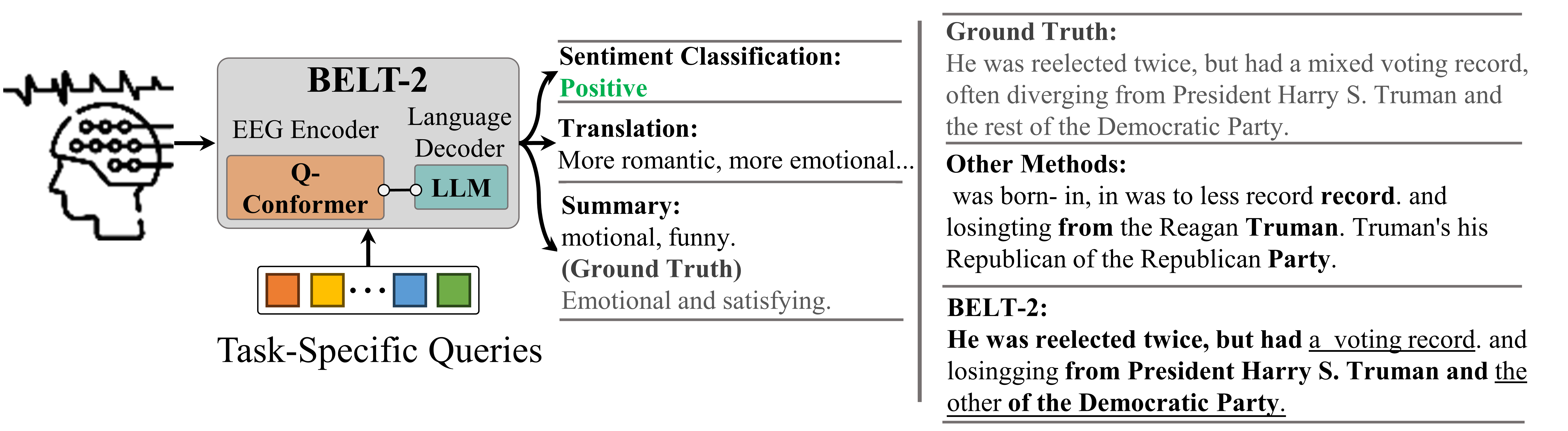}
    \caption{Overview of BELT-2. The first work of multi-task brain decoding by bridging the Q-Conformer EEG encoder and LLMs. Provided samples also suggest BELT-2 is the first to achieve fluent sentence decoding results from noninvasive brain signals.
 \label{fig:cover_main}}
\end{figure}

\section{Introduction}
Recently, the emergence of large language models (LLMs) has spurred efforts to integrate them with various modalities, such as VisualLLMs~\citep{liu2023visual,oquab2023dinov2}, and Robotics~\citep{driess2023palm}. These methods achieved remarkable improvement in various task settings. 
Yet, an important topic, the direct combination of LLMs with human intention remains relatively unexplored. 
Nonetheless, the inherent subject-wise non-stationary characteristics of Electroencephalography (EEG) signals, coupled with rigorous experimental protocols, make the task of decoding words or sentences exceptionally challenging.

Explorations on brain-to-text and brain-to-speech decoding in the earlier stage ~\citep{herff2015brain, makin2020machine, panachakel2021decoding,  nieto2021thinking} mostly perform decoding on a closed word-level set, which still has notable restrictions on vocabulary size and limitations to more intricate application scenarios. For the brain-to-language decoding, EEG-to-Text~\citep{wang2022open} introduced the open-vocabulary decoding of EEG signals with an initial performance baseline. 
DeWave~\citep{duan2023dewave} improved decoding performance by introducing a discrete encoder for EEG. BELT~\citep{zhou2023beltbootstrapping} which boosted decoding performance by leveraging language supervision. However, these methods are limited to single-task settings and have not achieved multi-task decoding from brain signals to natural languages. An extensive \textbf{related works} is provided in Appendix \ref{appendix:related-work} due to space limit. 

In this paper, we propose BELT-2, the first EEG-language learning framework to bridge the modality gap and effectively exploit LLM's generative capacity for EEG decoding. BELT-2 enhances three key aspects of brain decoding research.  
1) It is the first to introduce \textbf{BPE-level contrastive learning} for EEG-to-language alignment. 2) It first introduces a \textbf{prompt-based multi-task encoder} for EEG research. 3) It proposes a cost-effective solution for connecting an EEG encoder with a large language model (LLM).



More specifically, we introduce a novel discrete querying conformer (Q-Conformer) as the EEG encoder to improve encoding capacity and enable multitasking (Figure \ref{fig:QConformer}). Unlike previous single-task EEG encoders~\citep{zhou2023beltbootstrapping,duan2023dewave}, Q-Conformer is able to extract task-specific contexts according to a given query prompt. For the training of Q-Conformer, we propose the BPE-level EEG-language contrastive learning (BPE-CL) to bootstrap the learning of language-aligned EEG representation. After training, we bridge the Q-Conformer and an LLM decoder by prefix-tuning with both models frozen. To improve the performance of the briding, we further propose a technique called speculative augmentation (SA) to improve the training efficiency. The main contributions of BELT-2 could be concluded in four aspects.
\begin{itemize}
\setlength\itemsep{0mm}
\item 
This paper presents a novel framework capable of decoding fluent open-vocabulary sentences, facilitating multi-task EEG decoding including EEG translation, sentiment classification, and summarization.
\item 
The Q-Conformer is proposed to improve the encoding ability and the scalability for multi-tasking while the BPE-level contrastive learning establishes a firm alignment between EEG and language representations.
\item 
This paper provides a cost-effective bridging method for connecting LLMs with brain encodings by turning virtual-prefix. 
A speculative augmentation method is introduced to further improve the bridging performance. 
\item 
Experimental results suggest that the proposed BELT-2 exceeds SOTA performance on different EEG decoding tasks. For EEG translation, BELT-2 achieves $52.59$ BLEU-1, $17.85$ BLEU-4, and $40.1$ Rouge-1 Precision, which significantly outperforms the previous baseline by $31\%$, $162\%$ and $26\%$ respectively. On sentiment classification, BELT-2 achieves $74.62\%$ accuracy without further assistance from additional classifiers or external datasets. BELT-2 is also the first work that achieves EEG summarization with a SOTA $31.17$ BLEU-1 score.
\end{itemize}

\section{BELT-2}
BELT-2 introduces the Q-Conformer which enhances both the capacity to encode EEG information and the extendibility to multi-task. To bridge the modality gap between EEG and language, we boost EEG-to-Language representation learning through two learning stages: (1) the EEG-to-language alignment learning stage for learning the Q-Conformer EEG encoder. (2) a prefix-tuning stage for bridging Q-Conformer with LLM.

\subsection{Q-Conformer as EEG Encoder}
The overall structure of the Q-Conformer is illustrated in Figure \ref{fig:QConformer} which consists of a discrete conformer, a Context Transformer (C-Former), and a query prompt. The discrete conformer functions as a discrete EEG tokenizer that captures primitive patterns from the input EEG embeddings. The C-Former extracts mid-layer coding (MLC) that contains context information specific to a given task given by the learnable query prompt.  

\textbf{Discrete Conformer:}%
\quad
The discrete conformer consists of a conformer model and a vector quantizer. After preprocessing, the raw EEG waveform is segmented into windows using eye-tracking information. Then a frequency domain transform converts EEG segments into fix-size EEG embeddings $\mathbf{e}\in\mathbb{R}^{L\times{N}\times{D}}$. $L$ is the maximum length of the embedding sequence, $N$ denotes the number of EEG channels, and $D$ denotes the embedding size. The conformer model consists of $2$ conformer blocks which follow the structure manifested in \citep{gulati2020conformer}. The conformer model $E(\cdot)$ converts the EEG embeddings $\mathbf{e}$ into continuous EEG tokens $\mathbf{h}\in\mathbb{R}^{L\times{N}\times{d}}$, where $d$ denotes the size of the continuous EEG tokens.

We then convert $\mathbf{h}$ to a set of discrete tokens $\mathbf{b}$ by a vector quantizer (VQ) that looks up the nearest discrete code $\mathbf{v}_k, k=\{0,1,\cdots, K\}$ from the codebook $\mathcal{V}$~\citep{razavi2019generating}. The quantization process $\mathbf{z}_q(\mathbf{h})$ can be written as Equation \ref{eq:vq}. 
\begin{equation}\label{eq:vq}
  \mathbf{z}_q(\mathbf{h})=\{\mathbf{z}_q(\mathbf{h}_i)\}^L_{i=0}, \quad \mathbf{z}_q(\mathbf{h_i})=\mathbf{v}_k, \quad k =\argmin_j{\lVert\mathbf{h}_j-\mathbf{v}_j\rVert^2_2}
\end{equation}

We use $L_{vq}$ (Equation \ref{eq:vqloss}) to train the discrete codebook. The $L_{vq}$ is a weighted summation of $4$ loss terms. The first two terms are the codebook loss and the commitment loss. They are used to update the codebook by minimizing the information loss between the input and the output discrete tokens \cite{van2017neural}. The third term encourages the balanced use of all entries in the codebook and prevents codebook collapse during training \citep{dieleman2018challenge}. The last term is a reconstructive loss that ensures the information passed to the VQ is sufficient to describe the EEG signal.

\begin{equation}\label{eq:vqloss}
\mathcal{L}_{vq}=\lVert{sg}\left[\mathbf{h}\right]-\mathbf{z}_q(\mathbf{h})\rVert^2_2 + \lVert\mathbf{h}-sg\left[\mathbf{z}_q(\mathbf{h})\right] \rVert^2_2 + \frac{1}{\left|{\mathcal{V}}\right|}\sum^{\left|{\mathcal{V}}\right|}_{k=0}{p_k\log{p_k}} + \lVert{\mathbf{e}-\mathbf{\hat{e}}}\rVert^2_2
\end{equation}
, where $sg\left[\cdot\right]$ stands for the stop-gradient operator which is an identity at the forward pass while having zero gradients during the backward pass. $\left|{\mathcal{V}}\right|$ denotes the size of the discrete codebook and $p_k$ denotes the softmax probability of the codebook entry $k$ being used in each batch. $\mathbf{\hat{e}}$ denotes the reconstructed EEG embedding from $\mathbf{z}_q(\mathbf{h})$ using $2$ comformer blocks. 

\begin{figure}[!t]
    \centering
    \includegraphics[width=1.0\textwidth]{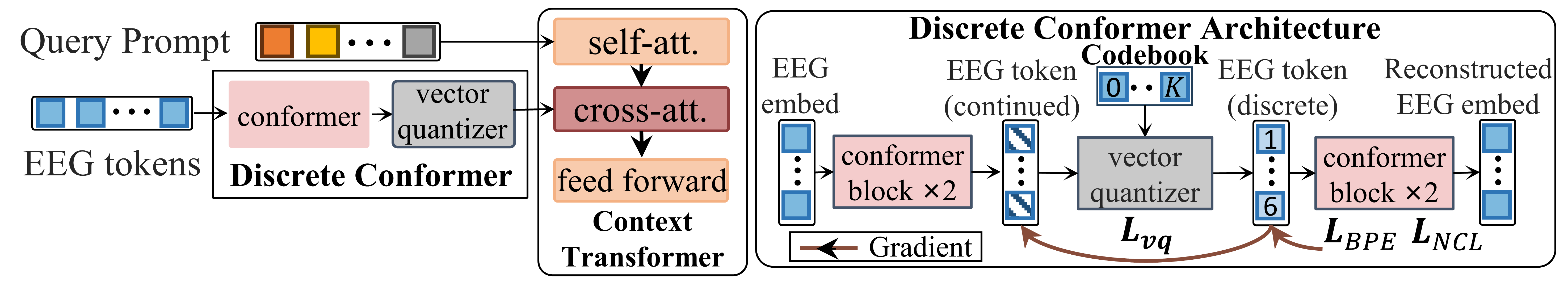}
    \caption{The overall structure of the Q-Conformer. It consists of a discrete conformer, a context transformer (C-Former), and a query prompt. The input EEG embeddings (EEG embed) are first processed by the conformer into continuous EEG tokens. A vector quantizer is then used to discretize the EEG tokens. Then, a query prompt interacts with the discrete EEG token via the cross-attention layer from in the C-Former to extract task-specific context information from the discrete EEG tokens.\label{fig:QConformer}}
\end{figure}

\textbf{C-Former and Query Prompt} 
\quad 
We create a set number of learnable query embeddings (query prompt) as input to the C-Former. The C-Former is composed of self-attention layers and cross-attention layers arranged in consecutive order. After feeding the query prompts and the discrete EEG tokens into the C-Former, the query prompts interact with each other through the self-attention layers and further interact with the discrete EEG tokens through the following cross-attention layer. A new query prompt will be initialized when training the Q-Conformer for a specific task. After training on a specific task, the query prompts learn to act as the instruction of the current task that guides the C-Former to extract MLC as the task-specific context from the EEG modality. 

This querying mechanism enables a more flexible adaptation of the pretrained Q-Conformer to a new downstream task by adding a new set of query prompts. It also allows the reusing of knowledge learned from previous training tasks. In our experiment setup, we initialize the C-Former with the pre-trained weights of  BART$_{large}$ \citep{lewis2019bart}. We employ a query prompt of $20$ learnable tokens for a specific, with each query possessing a dimensionality of $1024$.

\begin{figure*}[t]
    \centering
    \includegraphics[width=0.9\linewidth]{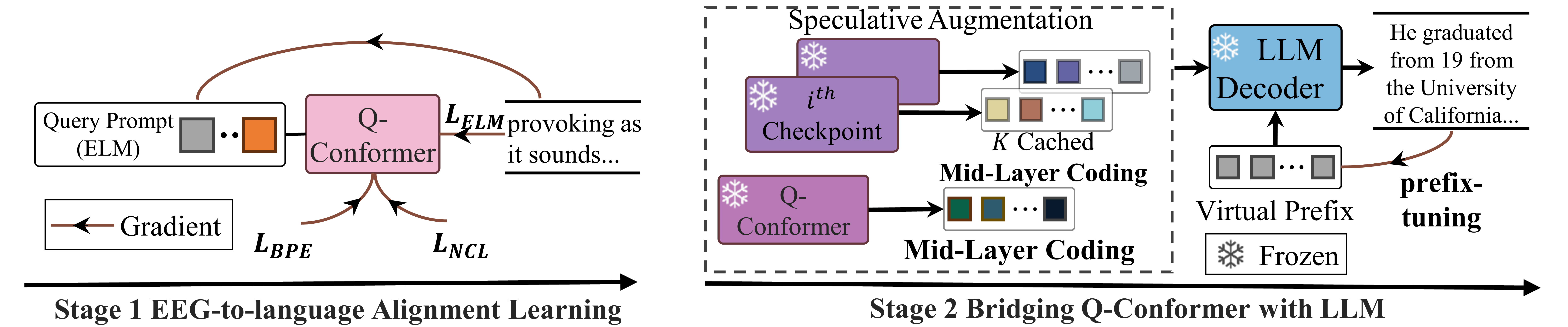}
    \caption{BELT-2's two-stage training schema. For EEG-to-language alignment learning (\textbf{left}), we jointly optimize three objectives that firmly establish the EEG-to-language alignment and enforce the query prompt to extract the EEG context most relevant to a task. For bridging of Q-Conformer and LLM (\textbf{right}), connect a frozen EEG model (Q-Conformer) and a frozen LLM by tuning the continuous virtual prefix using the prefix-tuning method. Speculative augmentation is used to boost the performance of the prefix-tuning process.}
    \label{fig:two-stage}\vspace{-1mm}
\end{figure*}

\subsection{EEG-to-language alignment learning}
In the EEG-to-language alignment learning stage, we train the Q-Conformer and align the encoded EEG tokens to the language modality. To achieve EEG-to-Language alignment, we combine two contrastive objectives and a pretraining objective to the VQ objective in Equation \ref{eq:vqloss}. The two contrastive objectives include (1) BPE-level contrastive learning (BPE-CL), and (2) Negative Contrastive learning (NCL). We further pretrain the Q-Conformer to achieve a task-specific query prompt by the EEG-to-Language matching (ELM) objective, which guides the C-Former to extract MLC that contains the most relevant EEG contexts in the specific task. 
\begin{wrapfigure}{r}{6cm}
\centering
\includegraphics[width=0.40\textwidth]{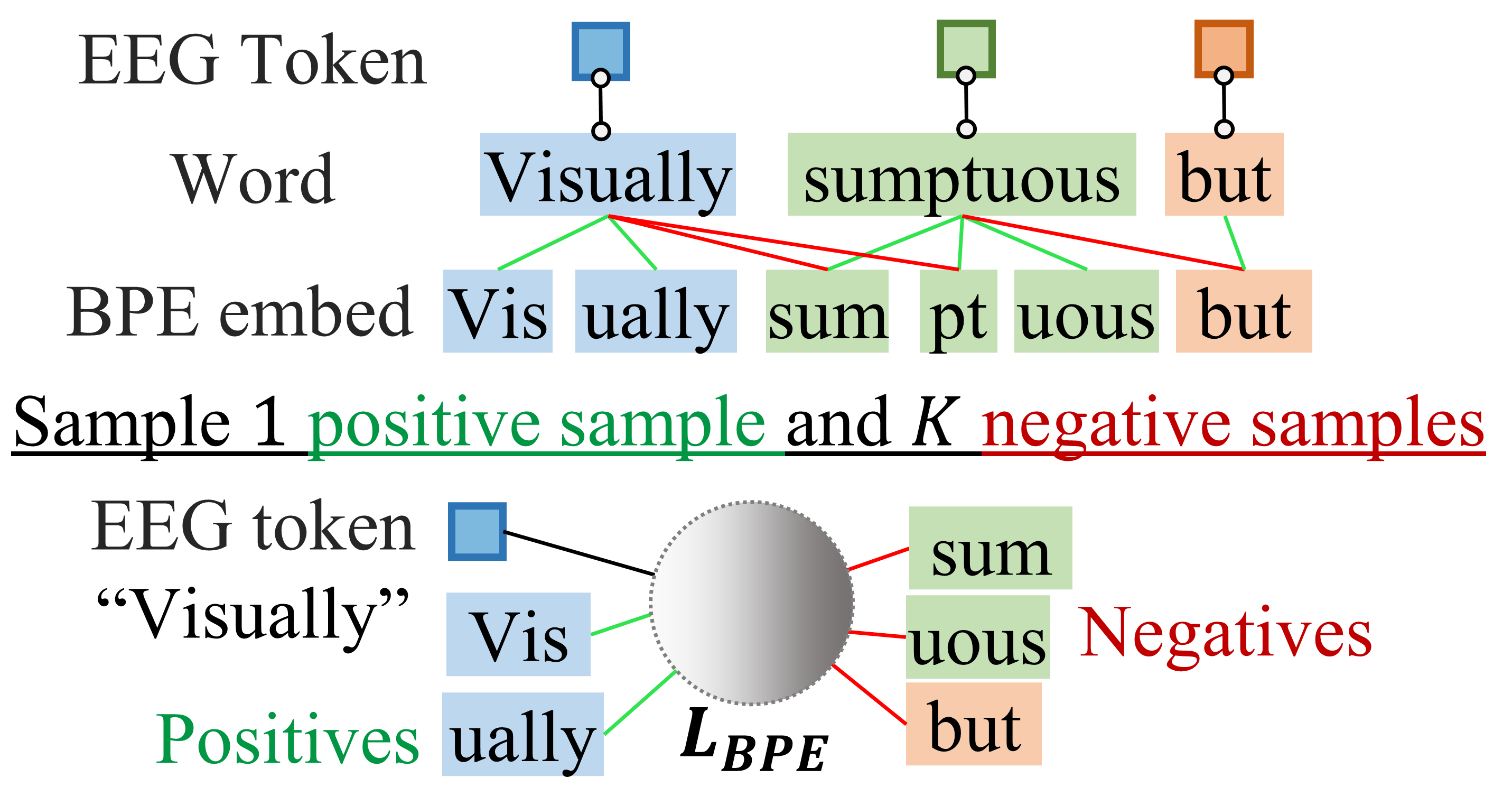}
\caption{\footnotesize The illustration of BPE-level contrastive learning.\label{fig:bpe-cl}}\vspace{-5mm}
\end{wrapfigure}

\textbf{BPE-level contrastive learning} (BPE-CL) learns to align the discrete EEG tokens with BPE subword embeddings by maximizing their mutual information. Unlike the BELT-1 model \citep{zhou2023beltbootstrapping} where contrastive learning is only performed at the word level, we perform EEG-language alignment in the BPE subword level to improve EEG-language alignment. Given the limited size of EEG-language pairs in the training set, this method enforces stronger semantic guidance to the EEG representation while enhancing the matching of subword units that are out-of-training vocabulary.  

The sampling strategy of the BPE-CL is illustrated in Figure \ref{fig:bpe-cl}. We commence by converting words into BPE tokens $\mathbf{w}\in\mathcal{W}$, e.g., converting $``Visually"$ into $[``Vis", ``ually"]$. The embeddings of these BPE tokens serve as positive targets for the EEG token corresponding to $``Visually"$ while BPE tokens other words are viewed as negative targets. We uniformly sample $1$ positive target and $K$ negative targets for each discrete EEG token in a training batch. The learning objective $L_{bpe}$ for the discrete EEG tokens and the BPE embeddings is formulated as:
\begin{equation} \label{eq:bpe-contrastive}
\mathcal{L}_{bpe}\!=-\log\frac{\exp{(\mathbf{z}_q(\mathbf{h})^\top{\mathbf{w}^+})}}{\exp{(\mathbf{z}_q(\mathbf{h})^\top{\mathbf{w}^+})}+\sum^K_{i=1}\exp{(\mathbf{z}_q(\mathbf{h})^\top{\mathbf{w}^-})}},
\end{equation}
, where $\mathbf{w}^+$ is the sampled embedding of the positive BPE token and $\mathbf{w}^-$ is the negative ones. 

\textbf{Negative contrastive learning} (NCL) aims to further improve the distinctions between the discrete EEG tokens by randomly sampling $K$ negative EEG tokens as distractors for each discrete EEG token in a training batch, which is defined as:
\begin{equation} \label{eq:negcontrastive}
\mathcal{L}_{neg}=-\log\frac{1}{\sum^K_{i=1}\exp{(\mathbf{z}_q(\mathbf{h})^\top{\mathbf{z}_q(\mathbf{h})^-})}},
\end{equation}
, where $\mathbf{z}_q(\mathbf{h})^-$ are sampled negative tokens from the batch and $\mathbf{z}_q(\mathbf{h})$ is defined in Equation \ref{eq:vq}. This objective enlarges the distinction among EEG tokens that are indistinguishable upon reading different words, easing the decoding effort.

\textbf{EEG-to-language matching} (ELM) aims to function as the pretraining task for learning the initial task-specific query prompt, which in terms is used to instruct the C-Former to extract task-specific context from the EEG tokens. We use a sequence-to-sequence machine translation loss similar to previous works \cite{zhou2023beltbootstrapping,wang2022open,duan2023dewave} as the objective function. Given the word-level EEG embedding sequence and text sentence pair $\left\langle\mathcal{E},\mathcal{S}\right\rangle$, we maximize the probability of the decoded sentence $p(\mathcal{S}|\mathcal{E})$ produced by the Q-Conformer. The learning objective is a machine translation term $L_{tr}$, which could be written as follows:
\begin{equation} \label{eq:loss_elm}
\mathcal{L}_{elm}=-\sum^L_l\log{p(s_l\in\mathcal{S}|\mathbf{q})}
\end{equation}
, where $L$ is the total length of the target text sequence, $s_l\in\mathcal{S}$ denotes the decoded tokens from the C-Former and $\mathbf{q}$ denotes the query prompt. 

\subsection{Bridging Q-Conformer with LLM}
We propose to bridge the frozen Q-Conformer and a frozen LLM to leverage both models effectively for EEG-to-Language tasks by tuning a set of virtual prefixes added to the output embeddings of the Q-Conformer, in order to achieve stronger performance at a lower training cost. 

\textbf{Prefix-tuning}\quad To achieve a proper prefix prompt that can steer the LLM to decode the MLC without changing the LLM's parameters, we adopt the prefix-tuning \citep{li2021prefix} method to only train a set of virtual prefix tokens as prompts to the LLM. In particular, we concat the virtual prefix and the MLC from the Q-Conformer as input to the subsequence frozen LLM. Please refer to Appendix \ref{sec:prefixtuning} for more details on prefix-tuning. 

\textbf{Speculative Augmentation} (SA) \quad
Despite the use of the lightweight prefix-tuning method, the size and diversity of training samples are still lacking. This is because while the Q-Conformer learns to extract task-specific context, it also learns to ignore task-irrelevant information. This would be a well-anticipated perk for an EEG encoder if we choose to directly decode language output from the EEG encoder. However, it also significantly reduces the diversity of training samples, making the learning of a good prefix difficult. 

Our BELT-2 framework solves this issues by proposing the SA method to sample MLC from a total of $K\!+\!1$ Q-Conformer checkpoints to provide more diverse prefix-tuning samples. In particular, we randomly sample $K$ model checkpoints other than the best-performing checkpoint to produce MLC for the prefix-tuning. During the forward process, a speculative ratio $r$ is defined to determine whether to use best checkpoint or one of the $K$ suboptimal checkpoints. To reduce the cost of memory, we cache the output MLCs of these $K$ model checkpoints during the training of Q-Conformer to avoid actually loading the checkpoints in the prefix-tuning stage. 

In our experiment, we set $K=15$ for a balance of performance and training costs to achieve a \textbf{$6\times$} larger and more diverse training sample set for the tuning of the LLM Decoder. 

\subsection{Extending Decoding to Multi-task}
\textbf{Translation:} \quad
Our definition of the EEG-to-Text translation task follows previous works on this topic~\citep{wang2022open}. Given the word-level EEG embedding sequence and text sentence pair $\left\langle\mathcal{E},\mathcal{S}\right\rangle$, we maximize the probability of the decoded sentence $p(\mathcal{S}|\mathcal{E})$ produced by our model. The training objective $L_{tr}$ for the translation task could be written as follows:
\begin{equation} \label{eq:tranlation-task-1}
p(\mathcal{S}|\mathcal{E}) = \prod^{L}_{l=1}p(s_l|\mathcal{E}, s_{<l}),\quad \mathcal{L}_{tr}=-\sum^L_l\log{p(s_l\in\mathcal{S})}
\end{equation}
where $L$ is the total length of the target text sequence and $s_l\in\mathcal{S}$ denotes the word tokens produced by our model.

\textbf{Summary:} \quad
We propose the first EEG-to-text summarization task by creating a summary dataset from the Zuco datasets. 
Human attention lingers around keywords and pivotal concepts during reading~\citep{ding2022cogbert}. Consequently, we hypothesize that the extraction of key concepts could be a more direct way to facilitate the transmission of neural information and the understanding of a person's intention. As such, our nuanced summarization task not only enhances our understanding of EEG data but also opens up exciting possibilities for advancing research in cognitive science. 

We kickstart by constructing the prompt "\textit{Rewrite the sentence by summarizing its main idea using $\{T\}$ words from the sentence, and keep the summarized sentence similar to the original sentence:$\{s\}$}" with $\{s\}$ being each ground truth sentence from the ZuCo dataset and attain the initial summarization targets for each sentence. We set $T\!=\!8$ in our experiment and use the LLAMA2 model \citep{touvron2023llama} to generate the initial summarization targets. Afterwards, manual inspection and rectification are carried out to improve the dataset's reliability and informativeness. The word-level EEG embedding sequence and summary pair are denoted by $\left\langle\mathcal{E},\hat{\mathcal{S}}\right\rangle$. To extend the Q-Conformer for summarization task, a new query prompt for summarization will be added. The training objective for generating summaries is similar to Equation \ref{eq:tranlation-task-1}, with the sole alteration being the substitution of $\mathcal{S}$ with $\hat{\mathcal{S}}$. For multi-task training, we train all tasks simultaneously by randomly sampling tasks for each update iteration.  

\begin{wrapfigure}{r}{6cm}
\centering
\includegraphics[width=0.40\textwidth]{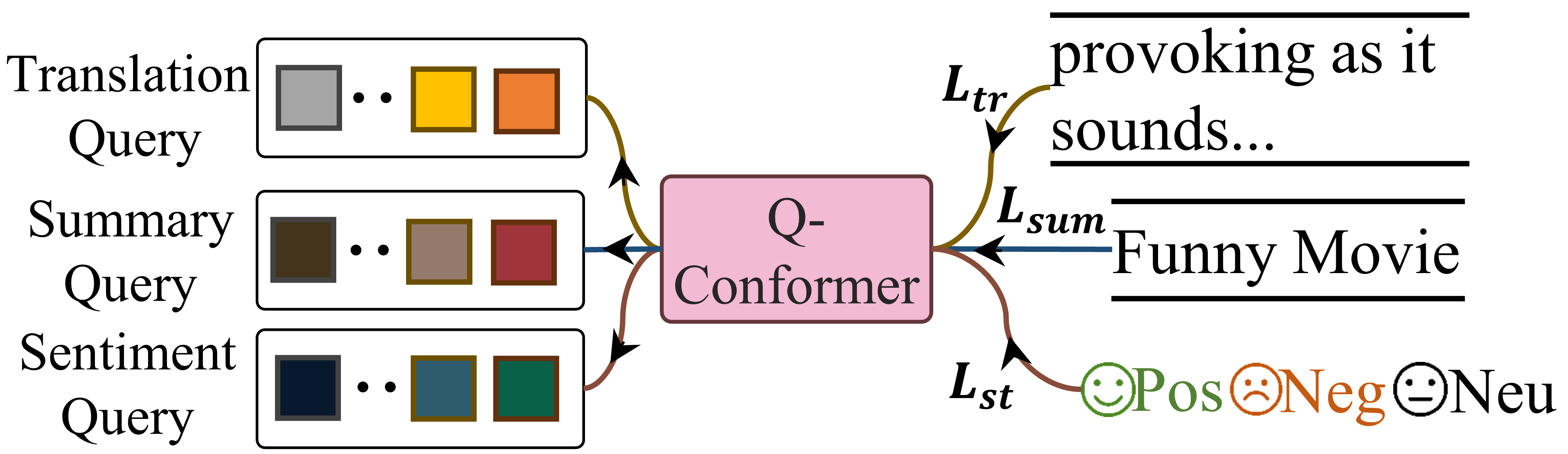}
\caption{\footnotesize For multi-task training, we train three tasks simultaneously by randomly sampling tasks for each training iteration. Each task-specific query prompt learns to provide task-specific instructions by training on the corresponding task-specific objective function. \label{fig:QConformer}}
\end{wrapfigure}

\textbf{Sentiment Classification:} \quad
We could further extend the Q-conformer to perform the sentiment classification task by adding another query prompt for the Q-Conformer and using the last output token from the Q-conformer as the CLS token. In particular, we use the EEG-sentiment label pair $\left\langle\mathcal{E},c\right\rangle$. Unlike \cite{wang2022open}, we don't need to use external sentiment classification datasets or learn an additional classifier. The training objective for sentiment classification is as follows:
\begin{equation} \label{eq:sst-task}
\mathcal{L}_{st} = -\sum^{\vert{C}\vert}\limits_{i=1}c_i\log p(\hat{c}|\mathcal{E}_i), 
\end{equation}
, where $\vert{C}\vert$ is the number of the sentiment categories and $\hat{c}$ is the sentiment prediction.

\section{Experiment and Results}
\subsection{Experiment Setup and Implementation Details}
We use the ZuCo datasets \citep{hollenstein2018zuco,hollenstein2019zuco} for the training and evaluation of the proposed BELT-2 framework. The ZuCo datasets contain EEG data recorded during natural reading tasks with eye-tracking data for word-level EEG segmentation. Reading material is collected from movie reviews \citep{socher2013recursive} and Wikipedia articles. We split the dataset into train, val, and test subsets (80\%,10\%,10\%). In this cross-sentence setting, sentences will not overlap among any two subsets. In addition, cross-subject performance is also evaluated. We evaluate translation and summary performance using the BLEU scores \citep{papineni2002bleu} and ROUGE-1 scores \cite{lin2004rouge}. We use \textbf{P.}, \textbf{R.}, \textbf{F1}, and \textbf{Acc.} to denote precision, recall, F1-score, and accuracy respectively. 

\subsection{Implementation Details}
The code could be assessed through an anonymous link~\footnote{https://anonymous.4open.science/r/BELT-2-0048}. For the word-level EEG embeddings, the total length of an embedding sequence is $L\!=\!56$ and the embedding size is $d\!=\!840$. The discrete conformer has $8$ attention heads with the feed-forward dimension size of $2048$ and a discrete codebook with $1024$ entries with a latent size of $1024$. The number of querying tokens used for The Q-Conformer is $20$. We train the Q-Conformer with a learning rate of $5e^{-06}$ for $60$ epochs during EEG-to-language alignment learning using AdamW \citep{loshchilov2017decoupled}. For the bridging stage, we use $8$ virtual prefix and set the speculative augmentation factor $K$ to $15$ with a speculative ratio of $0.3$. We use pre-trained BART and T5 models from the huggingface platform to initialize the Q-conformer and the LLM decoder. We also conducted experiments of massive size LLAMA2 model \footnote{https://huggingface.co/meta- llama/Llama-2-7b} in Section \ref{sec:ablation}. Due to the limitation of space, refer to Appendix \ref{appendix:implementation} for more details.

\subsection{Translation Performance}
\textbf{Quantitative Results}
We show quantitative results in Table \ref{tab:translation-result-main}. Compared to previous methods, e.g., EEG-to-Text \citep{wang2022open}, Dewave \citep{duan2023dewave}, and BELT-1 \citep{zhou2023beltbootstrapping} When only using EEG Encoder, We observe that the introduction of BPE-level contrastive learning bootstrapped a significant improvement (row $4$ compared to row $5$), achieving the SOTA EEG decoding BLEU-$\{1,2,3,4\}$ scores of $43.06,25.57,15.16,$ and $9.17$, which outperform DeWave by $1.71$, $1.42$, $1.24$, and $0.95$. By further connecting with the LLM decoder, BELT-2 further achieves the BLEU-$\{1,2,3,4\}$ scores of $52.59,36.32,25.21,$ and $17.85$, which brings additional $9.66$, $10.96$, $10.16$, and $8.76$ BLEU score improvements. The increase of the metrics is more significant for longer phrases ($+162\%$ for $4$-gram and $+99\%$ for $3$-gram) compared to the baseline EEG-to-Text method.
Additionally, we present ablation results that analyze the influence of VQ and the BPE-CL within our model, revealing that the utilization of BPE-CL significantly contributes to the enhancement of performance. However, multitask training did not bring a significant improvement to the translation result, which is elaborated in the Appendix \ref{Appendix:multitask-training}. 
\begin{table*}[h!]
\caption{Quantitative Results on Brain-to-Language Translation on the ZuCo Datasets.\label{tab:translation-result-main}}
\centering
\renewcommand\arraystretch{1}
\setlength\tabcolsep{2.6pt}
\resizebox{1.0\linewidth}{!}{
\begin{tabular}{ccccclllllll}
\toprule
\multirow{2}{*}{Model}                       & Vector                                & BPE-                                  & Enable                                & \multicolumn{1}{c}{Prefix}            & \multicolumn{4}{c}{\textbf{BLEU-N (\%)}}                                               & \multicolumn{3}{c}{\textbf{ROUGE-1 (\%)}}        \\ \cline{6-12} 
                                             & Quantizer                             & CL                                    & Multi-Task                            & \multicolumn{1}{c}{Tuning}            & \textbf{N=1}   & \textbf{N=2}   & \textbf{N=3}   & \multicolumn{1}{l|}{\textbf{N=4}}   & \textbf{R.}    & \textbf{P.}    & \textbf{F1}    \\ \hline
\multicolumn{1}{c|}{EEG-to-Text}             & \multicolumn{1}{c|}{$\times$}         & \multicolumn{1}{c|}{$\times$}         & \multicolumn{1}{c|}{$\times$}         & \multicolumn{1}{c|}{$\times$}          & 40.12          & 23.18          & 12.61          & \multicolumn{1}{l}{6.80}           & 28.8           & 31.7           & 30.1           \\
\multicolumn{1}{c|}{Dewave}                  & \multicolumn{1}{c|}{$\surd$}          & \multicolumn{1}{c|}{$\times$}         & \multicolumn{1}{c|}{$\times$}         & \multicolumn{1}{c|}{$\times$}          & 43.35          & 24.15          & 13.92          & \multicolumn{1}{l}{8.22}           & 28.82          & 33.71          & 30.69          \\ 
\multicolumn{1}{c|}{BELT-1}                  & \multicolumn{1}{c|}{$\surd$}          & \multicolumn{1}{c|}{$\times$}         & \multicolumn{1}{c|}{$\times$}         & \multicolumn{1}{c|}{$\times$}          & 42.31          & 25.26          & 14.81          & \multicolumn{1}{l}{8.73}           & 29.86          & 36.06          & 32.57          \\
\multicolumn{1}{c|}{\textbf{BELT-2}}         & \multicolumn{1}{c|}{\textbf{$\surd$}} & \multicolumn{1}{c|}{\textbf{$\surd$}} & \multicolumn{1}{c|}{\textbf{$\surd$}} & \multicolumn{1}{c|}{\textbf{$\times$}} & \textbf{43.06} & \textbf{25.57} & \textbf{15.05} & \multicolumn{1}{l}{\textbf{9.09}}  & \textbf{30.28} & \textbf{34.12} & \textbf{31.99} \\\hdashline
\multicolumn{1}{c|}{\textbf{BELT-2+LLM(T5)}} & \multicolumn{1}{c|}{\textbf{$\surd$}} & \multicolumn{1}{c|}{\textbf{$\surd$}} & \multicolumn{1}{c|}{\textbf{$\surd$}} & \multicolumn{1}{c|}{\textbf{$\surd$}}  & \textbf{52.38} & \textbf{36.28} & \textbf{25.28} & \multicolumn{1}{l}{\textbf{17.95}} & \textbf{36.08} & \textbf{39.47} & \textbf{37.59} \\ \hdashline
\multicolumn{12}{c}{\textbf{BELT-2 Ablations}}                                                                                                                                                                                                                                                                                                                     \\ \hdashline
\multicolumn{1}{c|}{BELT-2}                  & \multicolumn{1}{c|}{$\surd$}          & \multicolumn{1}{c|}{$\times$}         & \multicolumn{1}{c|}{$\surd$}          & \multicolumn{1}{c|}{$\times$}          & 41.57          & 24.02          & 13.80          & \multicolumn{1}{l}{8.06}           & 29.35          & 32.46          & 30.74          \\
\multicolumn{1}{c|}{BELT-2}                  & \multicolumn{1}{c|}{$\times$}         & \multicolumn{1}{c|}{$\surd$}          & \multicolumn{1}{c|}{$\surd$}          & \multicolumn{1}{c|}{$\times$}          & 41.90          & 24.57          & 14.2           & \multicolumn{1}{l}{8.28}           & 29.60          & 34.03          & 31.54          \\
\bottomrule
\end{tabular}
}
\end{table*}

\begin{table}[h!]
\renewcommand\arraystretch{1.1} 

\centering
\caption{Qualitative results on unseen EEG signals. The \textbf{bold} denotes an exact match between the ground truth and our prediction. \uline{underline} denotes a fuzzy match with similar semantic meanings.\label{tab:generation}}
\resizebox{0.96\textwidth}{!}{
\begin{tabular}{lll}
\toprule   
(1) & \multirow{3}{*}{Target} & \uline{\textbf{He} is a prominent \textbf{member} \textbf{of the} Bush \textbf{family}}, the \textbf{younger brother} \textbf{of President George} W. \textbf{Bush} \\
    &  &\textbf{and the} second son of former \textbf{President} \textbf{George H. W. Bush} and Barbara Bush. \\\cline{2-3} 
    & \multirow{3}{*}{Others} & was a former member of the American \textbf{family}, and first \textbf{brother} \textbf{of President George W. Bush.}                             \\
    &                         &  the father \textbf{son} of President \textbf{President George H. W. Bush}. his Bush.                         \\\cline{2-3}
    & \multirow{3}{*}{Ours}   & \uline{\textbf{He} was great member \textbf{member} \textbf{of the} American \textbf{family}}, and \textbf{younger brother}   \textbf{of President George} H. \textbf{Bush}                \\
    &                         &  \textbf{and the} younger cousin of President \textbf{President George H. W. Bush}. the Bush.                            \\\midrule
(2) & Target                  & \textbf{Adolf} Otto Reinhold Windaus \uline{(December 25, 1876 - June 9, 1959)} \textbf{was a} significant \textbf{German} chemist.                             \\ \cline{2-3} 
    & Others                  & rian Hitler,hardt,eren18 18, 1885 – January 3, 18) \textbf{was a German} figure- and                                                                       \\ \cline{2-3} 
    & Ours                    & \textbf{Adolf} Hitlero vonhard voner \uline{(J 15, 1875 - January 15, 1945)} \textbf{was a German} German industrialpacist                    \\ \midrule
(3) & Target                  & \textbf{It just doesn't} have   much else... especially \textbf{in} a moral \textbf{sense}.                                                       \\ \cline{2-3} 
    & Others                  & was so't work the to to and not the country \textbf{sense}.                                                                                       \\ \cline{2-3} 
    & Ours                    & \textbf{It just doesn't} work the of going except \textbf{in} the a way \textbf{sense}.                                                           \\ \midrule
(4) & \multirow{2}{*}{Target} & \textbf{He was reelected twice}, \uline{\textbf{but had} a mixed \textbf{voting record}}, often diverging \textbf{from}                       \\
    &                         & \textbf{President Harry S. Truman and the} \uline{rest of }\textbf{the Democratic Party}.                                                                 \\ \cline{2-3} 
    & \multirow{2}{*}{Others} & \textbf{was} a- in, in never to less record \textbf{record}. and losingting from his Reagan \textbf{Truman}.                                      \\
    &                         & Truman's his Republican of the Republican \textbf{Party}.                                                                                         \\ \cline{2-3} 
    & \multirow{2}{*}{Ours}   & \textbf{He was reelected twice}, \uline{\textbf{but had} voting \textbf{record}}. and   losingging \textbf{from}                              \\
    &                         & \textbf{President Harry S. Truman and the} \uline{other of} \textbf{the Democratic Party}.                                                    \\ \midrule
(5) & Target                  & Following the 1980 \textbf{presidential election}, Bush \textbf{and his family moved} \textbf{to Miami}-\textbf{Dade} \textbf{County, Florida.}   \\ \cline{2-3} 
    & Others                  & the deaths \textbf{election}, the was his wife \textbf{moved   to} California, \textbf{Dade County, Florida.}                                     \\ \cline{2-3} 
    & Ours                    & After his election \textbf{presidential election}, Reagan \textbf{and his family moved} \textbf{to Miami},\textbf{Dade} \textbf{County, Florida.} \\
\bottomrule
\end{tabular}}
\end{table}

\begin{figure}[h!]
    \centering
    \subfigure{
        \includegraphics[width=0.2\linewidth]{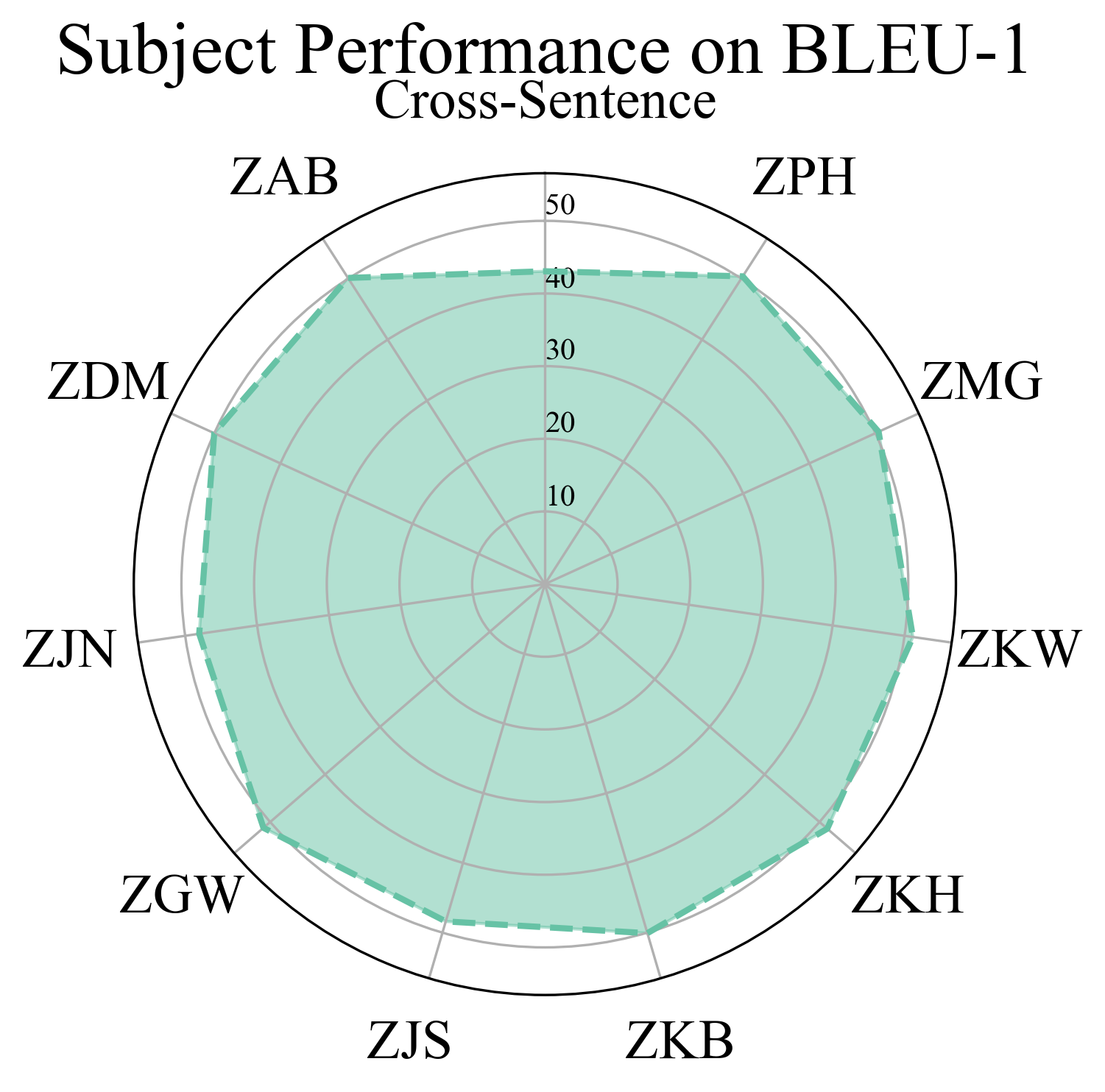}
    }\hspace{2mm}\vspace{0mm}
    \subfigure{
      \includegraphics[width=0.2\linewidth]{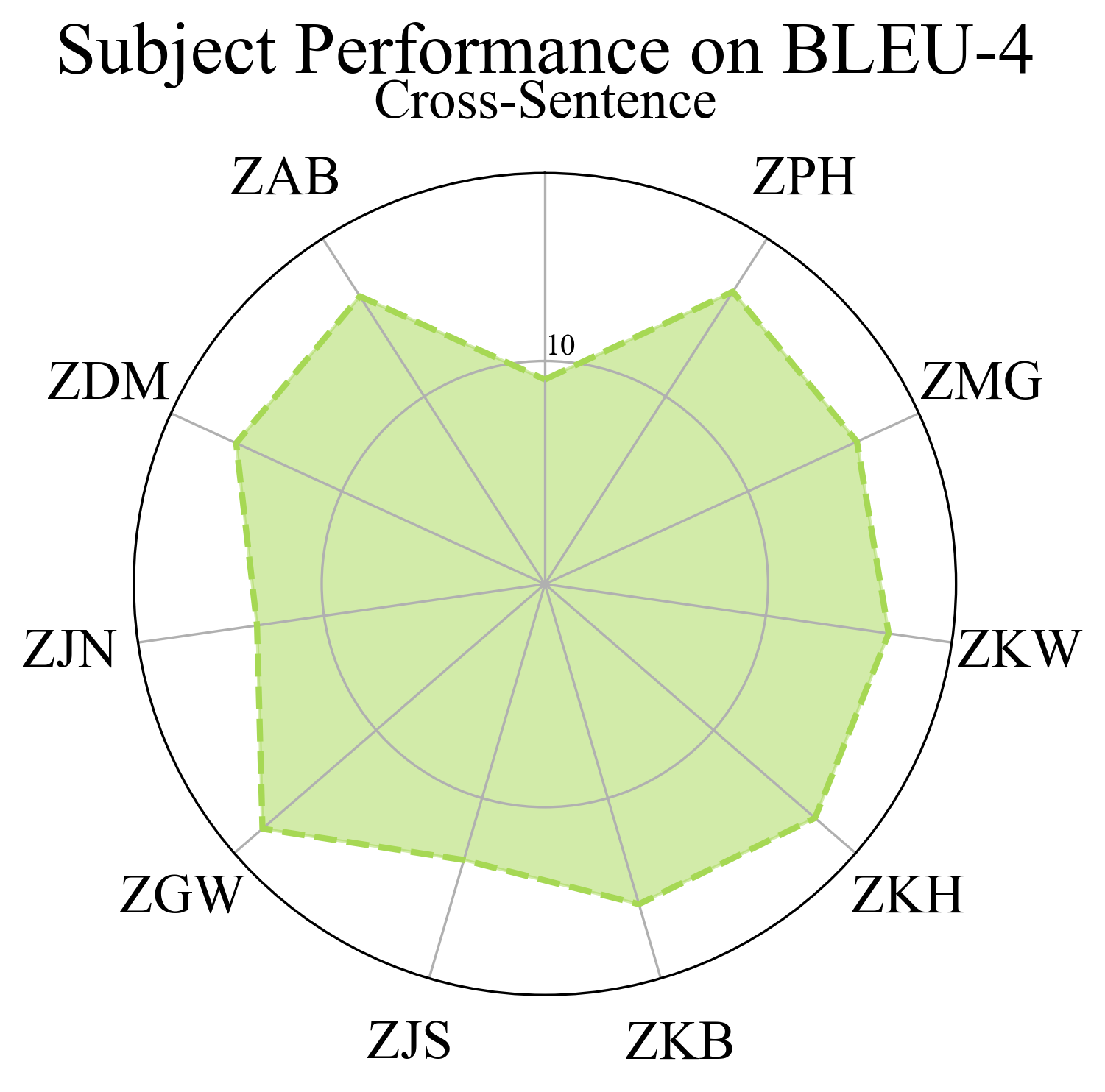}
    }\hspace{2mm}\vspace{0mm}
    \subfigure{
      \includegraphics[width=0.2\linewidth]{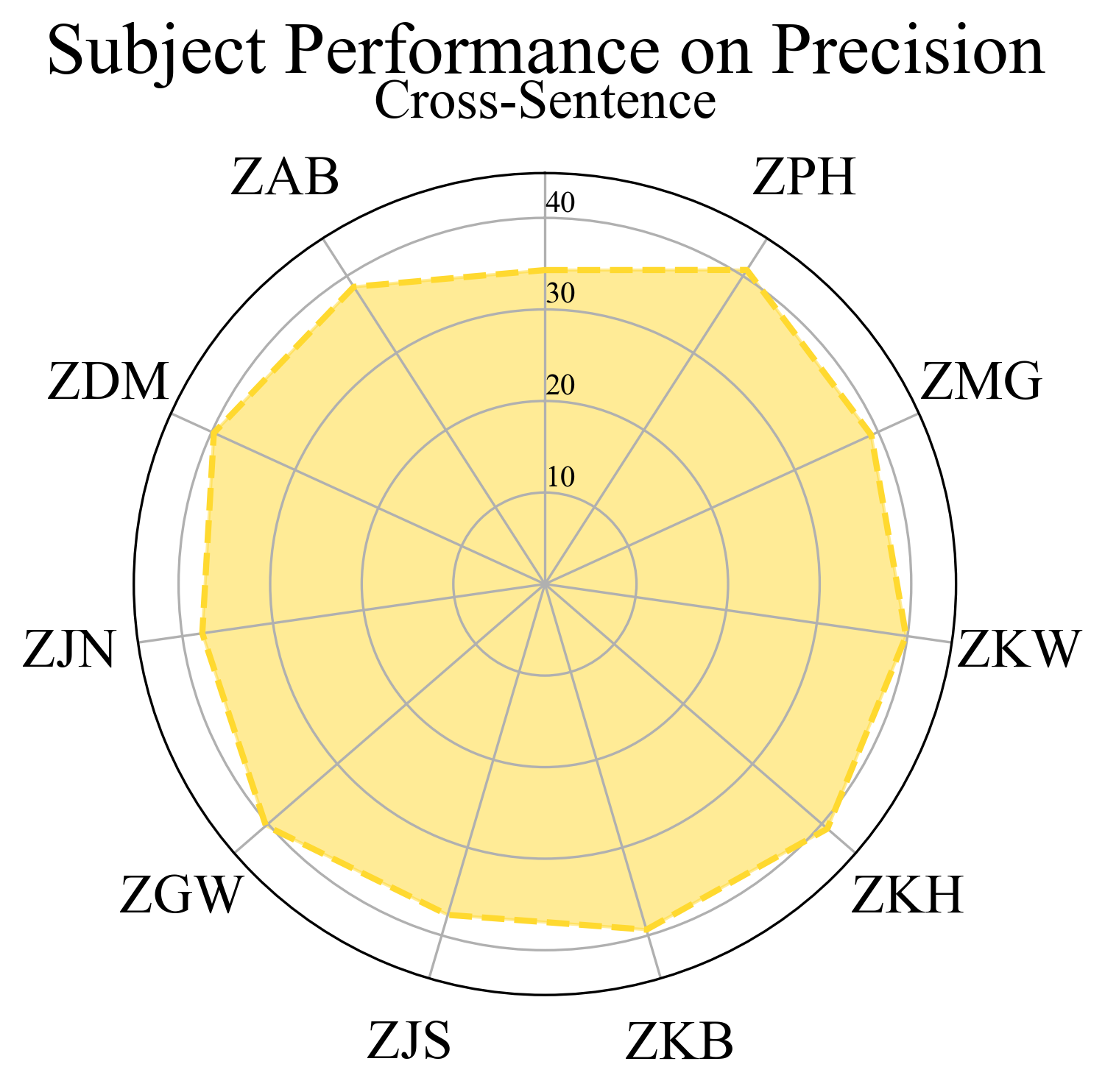}
    }\hspace{2mm}\vspace{0mm}
    \subfigure{
    \includegraphics[width=0.2\linewidth]{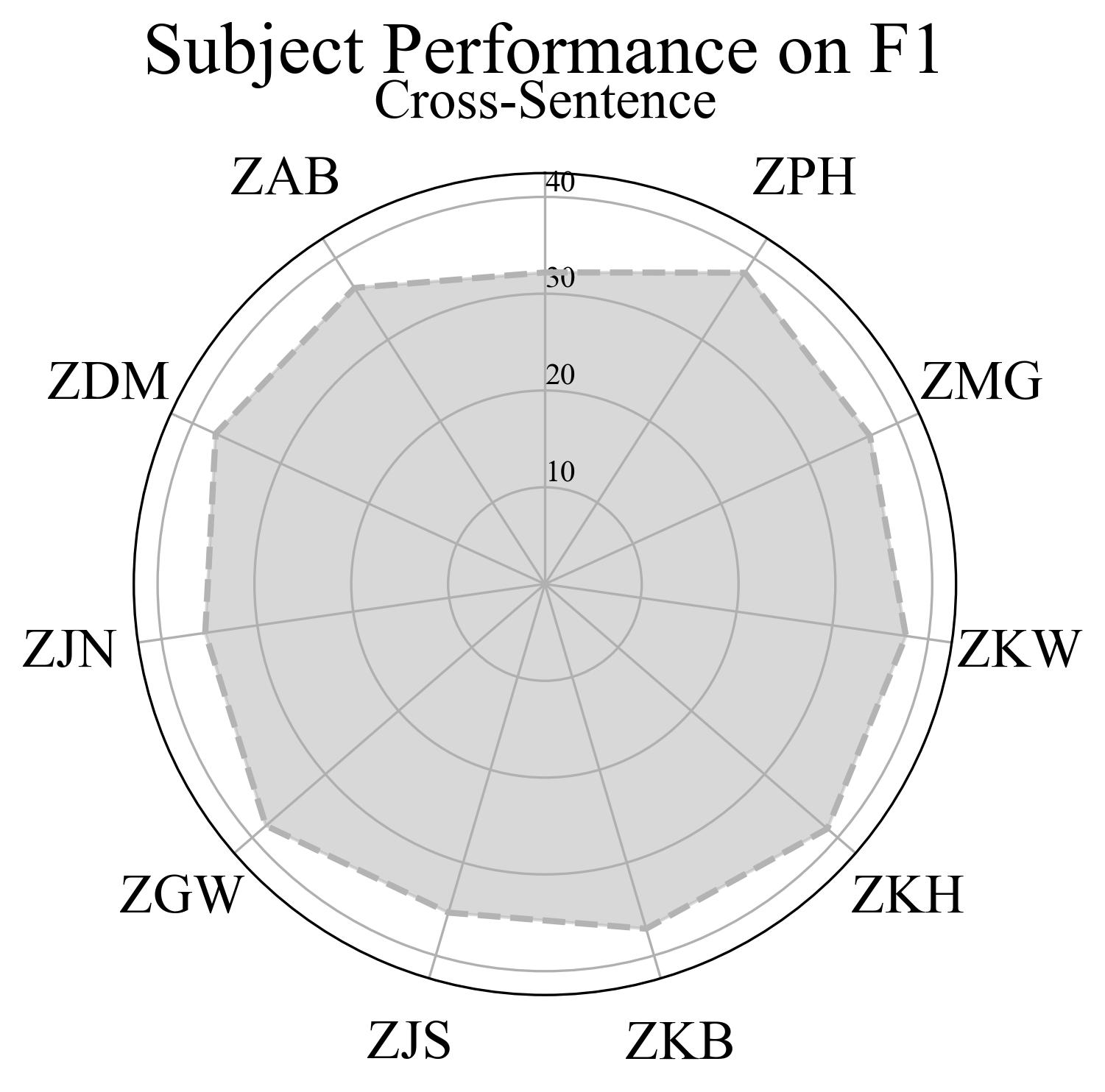}
    }\hspace{0mm}\vspace{0mm}
    \caption{The cross-subjects perfromance for translation task.}
    \label{fig:Cross-subject}
\end{figure}

\textbf{Cross-Subject Results}
As cross-subject performance is of vital importance for practical usage, we further report translation performance in cross-subject settings where we leave one subject out for evaluation and train the model using other subjects. Figure \ref{fig:Cross-subject} shows the cross-subject translation performance for a total of $10$ subjects compared to the cross-sentence result we achieved in the cross-sentence setting (Table \ref{tab:translation-result-main}). The radar charts in Figure \ref{fig:Cross-subject} denote the performance is stable across different subjects with subjects achieving BLEU-1 scores ranging from $48.04$ to $51.41$. 

\textbf{Qualitative Evaluation}
We showcase the generated text alongside the established approach from \cite{wang2022open} in Table \ref{tab:generation}. We observe that BELT-2 generates more fluent sentences with greater grammatical coherence. Notably, our model adeptly captures subject-predicate relationships while other methods miss the subject and predicate. This is demonstrated by the accurate decoding of phrases like ``\textit{He was}'' vs. ``\textit{He is}'', ``\textit{It just doesn't work}''  vs. ``\textit{It just doesn't have}''. Furthermore, for sentence structures involving quoted dates, such as ``\textit{(January 15, 1875 - January 15, 1945)}'' vs. ``\textit{(December 25, 1876 - June 9, 1959)}'', were also consistently deciphered. 

\subsection{Multi-task Performance}
\textbf{Sentiment Classification}
\quad
As shown in Table \ref{tab:sentiment-cls}, previous works need to train an LLM classifier using an external Stanford Sentiment Treebank dataset (around 11,000 sentences) \citep{socher2013recursive} and a new EEG encoder due to poor performance when training directly on the ZoCo dataset (Row $1$-$3$). In contrast, an EEG encoder incorporating external classifiers (row $4$-$7$) demonstrated improved performance \citep{wang2022open}. Our proposed Q-Conformer Encoder, achieve the state-of-the-art sentiment classification accuracy of $74.62\%$ on the ZuCo dataset. We also observe that our method could effectively leverage pretrained knowledge from the translation task to improve performance (row $8$-$9$).

\begin{figure}
\vspace{-0.2cm}
\begin{minipage}[b]{.38\linewidth}
    \centering
    \includegraphics[width=1.12\linewidth]{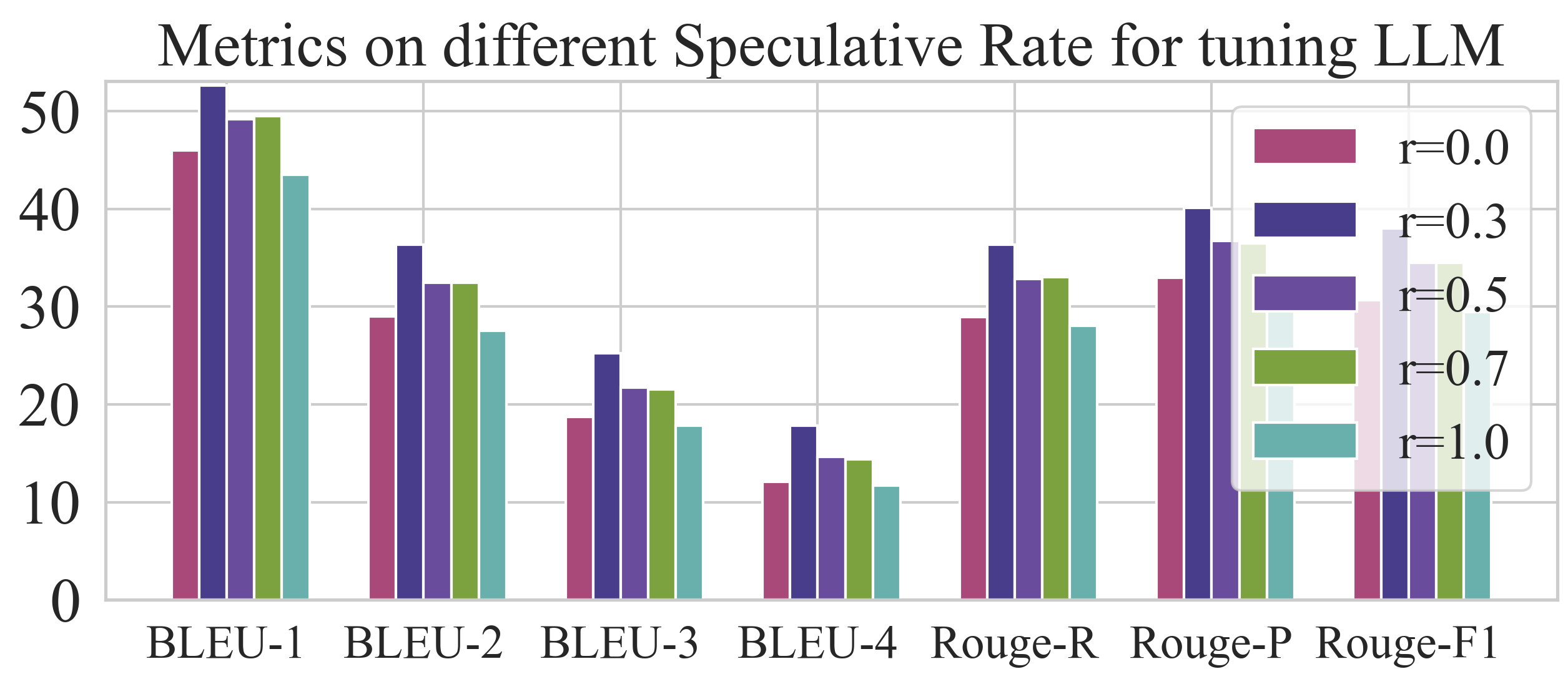}
    \caption{Ablation on speculative ratio. \label{fig:speculative-ratio}}
\end{minipage}
\raisebox{11pt}{
\begin{minipage}[b]{.62\linewidth}
\centering
\resizebox{0.85\linewidth}{!}{
\begin{tabular}{cccccc}
\toprule
\multirow{2}{*}{Model}    & \multicolumn{2}{c}{\textbf{BLEU (\%)}} & \multicolumn{3}{c}{\textbf{Rouge-1}} \\ \cline{2-6} 
                  & \textbf{N=1}                          & \textbf{N=3}                    & \textbf{P.}       & \textbf{R.}       & \textbf{F1} \\ \hline
EEG-to-Text               & 25.14                                 & 0                               & 10.37             & 7.30              & 8.49             \\
BELT-2 \textit{w/o Pretrained}   & 26.87                                 & 2.08                            & 9.84              & 11.06             & 10.34            \\
\textbf{BELT-2 \textit{w/ Pretrained}}   & \textbf{31.17}                        & \textbf{5.09}                   & \textbf{12.73}    & \textbf{13.26}    & \textbf{12.91}   \\
\bottomrule
\end{tabular}
}
\captionof{table}{Quantitative Results of Summary Task \label{tab:sum-result}}
\end{minipage}
}
\end{figure}

\textbf{Summarization}
\quad 
We compare the summarization performance of the BELT-2 model with the EEG-to-Text model as the baseline. As shown in Table \ref{tab:sum-result}, the EEG-to-Text struggles to generate summarization while the proposed BELT-2 model exhibited better generative capacity, especially in longer phrases. Compared to using a newly initialized encoder (row $2$), our BELT-2 exhibits a remarkable capacity to utilize the pretrained knowledge to increase the performance for the summarization task (row $3$). Generally, it attains the BLEU-$\{1,2,3,4\}$ scores of $31.17,15.7,8.91,5.09$, outperforming the baseline method.

\begin{table}[]
\caption{Quantitative Results of Sentiment Classification \label{tab:sentiment-cls}}
\centering
\resizebox{0.83\linewidth}{!}{
\begin{tabular}{ccccccc}
\toprule
\begin{tabular}[c]{@{}c@{}}EEG   \\ Encoder\end{tabular} & \begin{tabular}[c]{@{}c@{}}Additional \\ CLS Model\end{tabular} & \begin{tabular}[c]{@{}c@{}}Additional \\ Dataset \end{tabular}  & \textbf{Acc.}  & \textbf{P.} & \textbf{R.} & \textbf{F1}    \\ \hline
MLP   & \multirow{1}{*}{None}       & \multirow{1}{*}{None}   & \multirow{1}{*}{31.8}   & \multirow{1}{*}{32.8}  & \multirow{1}{*}{33.6}  & \multirow{1}{*}{27.5} \\
Bi-LSTM & \multirow{1}{*}{None}&\multirow{1}{*}{None}&\multirow{1}{*}{30.9}&\multirow{1}{*}{27.5}&\multirow{1}{*}{33.6}& \multirow{1}{*}{17.4} \\
Transformer          & \multirow{1}{*}{BERT}& \multirow{1}{*}{None}&\multirow{1}{*}{36.6}   &\multirow{1}{*}{23.7}   &\multirow{1}{*}{34.5}   &\multirow{1}{*}{27.2}  \\
EEG2Text   & \multirow{1}{*}{BART}   & \multirow{1}{*}{SST}& \multirow{1}{*}{55.30}           & \multirow{1}{*}{62.40}     & \multirow{1}{*}{56.50}  & \multirow{1}{*}{55.60} \\
BELT-1                                                   & BART                                                            & SST         & 65.13          & 63.67     & 63.34  & 62.45 \\
BELT-1                                                   & Albertv2                                                        & SST         & 60.09          & 61.63     & 60.03  & 59.56 \\
BELT-1                                                   & XLNet                                                           & SST         & 67.32          & 66.55     & 65.71  & 65.02 \\
BELT-2 \textit{w/o Pretrained}  & None & None    & 59.74 & 57.67 &57.63 &57.11  \\
\textbf{BELT-2 \textit{w/ Pretrained}}& \textbf{None}&\textbf{None}&\textbf{74.62}&\textbf{75.34} &\textbf{73.84}& \textbf{73.31}\\
\bottomrule
\end{tabular}}
\end{table}

\subsection{Ablation Study}\label{sec:ablation}

\textbf{Bridging Q-Confomer Encoder with different LLMs}
\quad
Table \ref{tab:translation-result-main} shows the result of bridging our Q-Conformer encoder with the T5 \citep{raffel2020exploring}. In Table \ref{tab:translation-stage2-ablation}, we conduct a comprehensive investigation of bridging LLM decoders with the Q-Conformer model, including the LLAMA2, T5, and the PEGASUS \citep{zhang2020pegasus} models. Results show that T5 LLMs consistently outperform other variants and boost the decoding performance. We attribute this superiority to T5's denoising training objectives. However, the sheer scale of the LLM decoder does not necessarily lead to enhanced decoding performance. For example, PEGASUS and LLAMA2 did not yield much improvement in the translation performance. 


\begin{table*}[h!]
\caption{Ablation study of bridging Q-Confomer Encoder with different LLMs \label{tab:translation-stage2-ablation}}
\centering
\resizebox{0.85\linewidth}{!}{
\setlength\tabcolsep{4pt}
\begin{tabular}{lllllllll}
\toprule
              &                       & \multicolumn{4}{c}{\textbf{BLEU-N (\%)}}                                         & \multicolumn{3}{c}{\textbf{ROUGE-1 (\%)} }                      \\\cline{3-9}  
\textbf{LLM}       & \textbf{Type}                 & \textbf{N=1}    & \textbf{N=2 }  & \textbf{N=3}   & \multicolumn{1}{l}{\textbf{N=4}}   & \textbf{P.}     & \textbf{R.}     & \textbf{F1 }   \\  \midrule
LLAMA2  & 7B                    & 21.40          & 6.96           & 3.38           & 2.21           & 12.23          & 13.20          & 12.61          \\ \midrule
PEGASUS & google/pegasus-x-base & 37.67          & 18.90          & 9.68           & 5.21           & 26.43          & 31.06          & 28.38          \\
              & google/pegasus-xsum   & 40.82          & 23.70          & 13.39          & 7.61           & 30.25          & 33.94          & 31.86          \\ \hline
T5      & t5-small              & 51.02          & 33.44          & 22.41          & 15.42          & 34.91          & 37.80     & 36.15          \\
          & t5-base               & 51.36          & 33.75          & 22.74          & 15.63          & 35.09          & 38.19          & 36.41          \\
          & \textbf{t5-large}     & \textbf{52.59} & \textbf{36.32} & \textbf{25.21} & \textbf{17.85} & \textbf{36.32} & \textbf{40.10} & \textbf{38.00} \\
          & google/flan-t5-base   & 50.01          & 33.09          & 21.77          & 14.49          & 32.97          & 36.64          & 34.54          \\
          & google/flan-t5-large  & 49.85          & 33.08          & 22.07          & 14.84          & 33.11          & 36.61          & 34.59          \\
\bottomrule
\end{tabular}
}
\end{table*}

\textbf{Speculative Augmentation}\quad
We further conduct ablation experiments on the effect of different speculative ratios in Figure \ref{fig:speculative-ratio}. We observe that the introduction of speculative augmentation at $r\!=\!0.3$ has a significantly better impact on the decoding performance across all evaluated metrics. 

\subsection*{\textbf{Limitations}}
While BELT-2 achieved remarkable translation improvements by combining Q-Conformer with LLMs, it is worth noting that the accuracy still lags behind traditional language-to-language translation. 
Also, it is noted that the experiments were conducted on publicly available neural reading datasets with the help of eye-tracking markers. As a result, BELT-2 has not realized everyday communication such as `silent speech' or `reading mind'. The vision of communication or controlling devices directly from brain dynamics remains a challenging task for follow-up research. 

\section{Conclusion}
This paper introduces BELT-2, a pioneering EEG-language learning framework for bridging brain signals to LLMs. Our framework achieves EEG-to-language alignment by incorporating the novel BPE-CL objective and proposed an effective method for bridging a frozen Q-Conformer EEG Encoder and a frozen LLM to leverage their generative capacity. The multi-task extendibility of the Q-Conformer also establishes BELT-2 as the first work to achieve a multi-task decoding model in EEG research. Extensive experiments were conducted to evaluate the performance of BELT-2 quantitatively and qualitatively. Especially, this work provides the first study investigating the feasibility of using frozen pretrained LLM to process EEG contexts exampled by a wide range of LLMs. Our experimental result shows that the BELT-2 framework represents a significant step forward in integrating human brain signals with LLMs, opening up exciting new avenues for research and development in cognitive neuroscience and brain-computer interfaces. We hope that this work will inspire further exploration and innovation in this exciting and rapidly evolving field.



\bibliography{iclr2024_conference}

\begin{thebibliography}{42}
\providecommand{\natexlab}[1]{#1}
\providecommand{\url}[1]{\texttt{#1}}
\expandafter\ifx\csname urlstyle\endcsname\relax
  \providecommand{\doi}[1]{doi: #1}\else
  \providecommand{\doi}{doi: \begingroup \urlstyle{rm}\Url}\fi

\bibitem[Anumanchipalli et~al.(2019)Anumanchipalli, Chartier, and
  Chang]{anumanchipalli2019speech}
Gopala~K Anumanchipalli, Josh Chartier, and Edward~F Chang.
\newblock Speech synthesis from neural decoding of spoken sentences.
\newblock \emph{Nature}, 568\penalty0 (7753):\penalty0 493--498, 2019.

\bibitem[Cruttenden(2014)]{cruttenden2014gimson}
Alan Cruttenden.
\newblock \emph{Gimson's pronunciation of English}.
\newblock Routledge, 2014.

\bibitem[Desai \& Johnson(2021)Desai and Johnson]{desai2021virtex}
Karan Desai and Justin Johnson.
\newblock Virtex: Learning visual representations from textual annotations.
\newblock In \emph{Proceedings of the IEEE/CVF conference on computer vision
  and pattern recognition}, pp.\  11162--11173, 2021.

\bibitem[Dieleman et~al.(2018)Dieleman, van~den Oord, and
  Simonyan]{dieleman2018challenge}
Sander Dieleman, Aaron van~den Oord, and Karen Simonyan.
\newblock The challenge of realistic music generation: modelling raw audio at
  scale.
\newblock \emph{Advances in neural information processing systems}, 31, 2018.

\bibitem[Ding et~al.(2022)Ding, Chen, Du, Qin, and Liu]{ding2022cogbert}
Xiao Ding, Bowen Chen, Li~Du, Bing Qin, and Ting Liu.
\newblock Cogbert: Cognition-guided pre-trained language models.
\newblock In \emph{Proceedings of the 29th International Conference on
  Computational Linguistics}, pp.\  3210--3225, 2022.

\bibitem[Dosovitskiy et~al.(2020)Dosovitskiy, Beyer, Kolesnikov, Weissenborn,
  Zhai, Unterthiner, Dehghani, Minderer, Heigold, Gelly,
  et~al.]{dosovitskiy2020image}
Alexey Dosovitskiy, Lucas Beyer, Alexander Kolesnikov, Dirk Weissenborn,
  Xiaohua Zhai, Thomas Unterthiner, Mostafa Dehghani, Matthias Minderer, Georg
  Heigold, Sylvain Gelly, et~al.
\newblock An image is worth 16x16 words: Transformers for image recognition at
  scale.
\newblock \emph{arXiv preprint arXiv:2010.11929}, 2020.

\bibitem[Driess et~al.(2023)Driess, Xia, Sajjadi, Lynch, Chowdhery, Ichter,
  Wahid, Tompson, Vuong, Yu, et~al.]{driess2023palm}
Danny Driess, Fei Xia, Mehdi~SM Sajjadi, Corey Lynch, Aakanksha Chowdhery,
  Brian Ichter, Ayzaan Wahid, Jonathan Tompson, Quan Vuong, Tianhe Yu, et~al.
\newblock Palm-e: An embodied multimodal language model.
\newblock \emph{arXiv preprint arXiv:2303.03378}, 2023.

\bibitem[Duan et~al.(2023)Duan, Zhou, Wang, Wang, and Lin]{duan2023dewave}
Yiqun Duan, Jinzhao Zhou, Zhen Wang, Yu-Kai Wang, and Chin-Teng Lin.
\newblock Dewave: Discrete eeg waves encoding for brain dynamics to text
  translation.
\newblock \emph{arXiv preprint arXiv:2309.14030}, 2023.

\bibitem[Elizalde et~al.(2023)Elizalde, Deshmukh, Al~Ismail, and
  Wang]{elizalde2023clap}
Benjamin Elizalde, Soham Deshmukh, Mahmoud Al~Ismail, and Huaming Wang.
\newblock Clap learning audio concepts from natural language supervision.
\newblock In \emph{ICASSP 2023-2023 IEEE International Conference on Acoustics,
  Speech and Signal Processing (ICASSP)}, pp.\  1--5. IEEE, 2023.

\bibitem[Gulati et~al.(2020)Gulati, Qin, Chiu, Parmar, Zhang, Yu, Han, Wang,
  Zhang, Wu, et~al.]{gulati2020conformer}
Anmol Gulati, James Qin, Chung-Cheng Chiu, Niki Parmar, Yu~Zhang, Jiahui Yu,
  Wei Han, Shibo Wang, Zhengdong Zhang, Yonghui Wu, et~al.
\newblock Conformer: Convolution-augmented transformer for speech recognition.
\newblock \emph{arXiv preprint arXiv:2005.08100}, 2020.

\bibitem[Herff et~al.(2015)Herff, Heger, De~Pesters, Telaar, Brunner, Schalk,
  and Schultz]{herff2015brain}
Christian Herff, Dominic Heger, Adriana De~Pesters, Dominic Telaar, Peter
  Brunner, Gerwin Schalk, and Tanja Schultz.
\newblock Brain-to-text: decoding spoken phrases from phone representations in
  the brain.
\newblock \emph{Frontiers in neuroscience}, 9:\penalty0 217, 2015.

\bibitem[Hollenstein et~al.(2018)Hollenstein, Rotsztejn, Troendle, Pedroni,
  Zhang, and Langer]{hollenstein2018zuco}
Nora Hollenstein, Jonathan Rotsztejn, Marius Troendle, Andreas Pedroni,
  Ce~Zhang, and Nicolas Langer.
\newblock Zuco, a simultaneous eeg and eye-tracking resource for natural
  sentence reading.
\newblock \emph{Scientific data}, 5\penalty0 (1):\penalty0 1--13, 2018.

\bibitem[Hollenstein et~al.(2019)Hollenstein, Troendle, Zhang, and
  Langer]{hollenstein2019zuco}
Nora Hollenstein, Marius Troendle, Ce~Zhang, and Nicolas Langer.
\newblock Zuco 2.0: A dataset of physiological recordings during natural
  reading and annotation.
\newblock \emph{arXiv preprint arXiv:1912.00903}, 2019.

\bibitem[Joulin et~al.(2016)Joulin, Van Der~Maaten, Jabri, and
  Vasilache]{joulin2016learning}
Armand Joulin, Laurens Van Der~Maaten, Allan Jabri, and Nicolas Vasilache.
\newblock Learning visual features from large weakly supervised data.
\newblock In \emph{Computer Vision--ECCV 2016: 14th European Conference,
  Amsterdam, The Netherlands, October 11--14, 2016, Proceedings, Part VII 14},
  pp.\  67--84. Springer, 2016.

\bibitem[Lewis et~al.(2019)Lewis, Liu, Goyal, Ghazvininejad, Mohamed, Levy,
  Stoyanov, and Zettlemoyer]{lewis2019bart}
Mike Lewis, Yinhan Liu, Naman Goyal, Marjan Ghazvininejad, Abdelrahman Mohamed,
  Omer Levy, Ves Stoyanov, and Luke Zettlemoyer.
\newblock Bart: Denoising sequence-to-sequence pre-training for natural
  language generation, translation, and comprehension.
\newblock \emph{arXiv preprint arXiv:1910.13461}, 2019.

\bibitem[Li et~al.(2023)Li, Li, Savarese, and Hoi]{li2023blip}
Junnan Li, Dongxu Li, Silvio Savarese, and Steven Hoi.
\newblock Blip-2: Bootstrapping language-image pre-training with frozen image
  encoders and large language models.
\newblock \emph{arXiv preprint arXiv:2301.12597}, 2023.

\bibitem[Li \& Liang(2021)Li and Liang]{li2021prefix}
Xiang~Lisa Li and Percy Liang.
\newblock Prefix-tuning: Optimizing continuous prompts for generation.
\newblock \emph{arXiv preprint arXiv:2101.00190}, 2021.

\bibitem[Lin(2004)]{lin2004rouge}
Chin-Yew Lin.
\newblock Rouge: A package for automatic evaluation of summaries.
\newblock In \emph{Text summarization branches out}, pp.\  74--81, 2004.

\bibitem[Liu et~al.(2023{\natexlab{a}})Liu, Chen, Yuan, Mei, Liu, Mandic, Wang,
  and Plumbley]{liu2023audioldm}
Haohe Liu, Zehua Chen, Yi~Yuan, Xinhao Mei, Xubo Liu, Danilo Mandic, Wenwu
  Wang, and Mark~D Plumbley.
\newblock Audioldm: Text-to-audio generation with latent diffusion models.
\newblock \emph{arXiv preprint arXiv:2301.12503}, 2023{\natexlab{a}}.

\bibitem[Liu et~al.(2023{\natexlab{b}})Liu, Li, Wu, and Lee]{liu2023visual}
Haotian Liu, Chunyuan Li, Qingyang Wu, and Yong~Jae Lee.
\newblock Visual instruction tuning.
\newblock \emph{arXiv preprint arXiv:2304.08485}, 2023{\natexlab{b}}.

\bibitem[Loshchilov \& Hutter(2017)Loshchilov and
  Hutter]{loshchilov2017decoupled}
Ilya Loshchilov and Frank Hutter.
\newblock Decoupled weight decay regularization.
\newblock \emph{arXiv preprint arXiv:1711.05101}, 2017.

\bibitem[Makin et~al.(2020)Makin, Moses, and Chang]{makin2020machine}
Joseph~G Makin, David~A Moses, and Edward~F Chang.
\newblock Machine translation of cortical activity to text with an
  encoder--decoder framework.
\newblock \emph{Nature neuroscience}, 23\penalty0 (4):\penalty0 575--582, 2020.

\bibitem[Mokady et~al.(2021)Mokady, Hertz, and Bermano]{mokady2021clipcap}
Ron Mokady, Amir Hertz, and Amit~H Bermano.
\newblock Clipcap: Clip prefix for image captioning.
\newblock \emph{arXiv preprint arXiv:2111.09734}, 2021.

\bibitem[Moses et~al.(2021)Moses, Metzger, Liu, Anumanchipalli, Makin, Sun,
  Chartier, Dougherty, Liu, Abrams, Tu-Chan, Ganguly, and
  Chang]{doi:10.1056/NEJMoa2027540}
David~A. Moses, Sean~L. Metzger, Jessie~R. Liu, Gopala~K. Anumanchipalli,
  Joseph~G. Makin, Pengfei~F. Sun, Josh Chartier, Maximilian~E. Dougherty,
  Patricia~M. Liu, Gary~M. Abrams, Adelyn Tu-Chan, Karunesh Ganguly, and
  Edward~F. Chang.
\newblock Neuroprosthesis for decoding speech in a paralyzed person with
  anarthria.
\newblock \emph{New England Journal of Medicine}, 385\penalty0 (3):\penalty0
  217--227, 2021.
\newblock \doi{10.1056/NEJMoa2027540}.
\newblock URL \url{https://doi.org/10.1056/NEJMoa2027540}.

\bibitem[Nieto et~al.(2021)Nieto, Peterson, Rufiner, Kamienkowski, and
  Spies]{nieto2021thinking}
Nicolas Nieto, Victoria Peterson, Hugo~Leonardo Rufiner, Juan Kamienkowski, and
  Ruben Spies.
\newblock " thinking out loud": an open-access eeg-based bci dataset for inner
  speech recognition.
\newblock \emph{bioRxiv}, 2021.

\bibitem[Oquab et~al.(2023)Oquab, Darcet, Moutakanni, Vo, Szafraniec, Khalidov,
  Fernandez, Haziza, Massa, El-Nouby, et~al.]{oquab2023dinov2}
Maxime Oquab, Timoth{\'e}e Darcet, Th{\'e}o Moutakanni, Huy Vo, Marc
  Szafraniec, Vasil Khalidov, Pierre Fernandez, Daniel Haziza, Francisco Massa,
  Alaaeldin El-Nouby, et~al.
\newblock Dinov2: Learning robust visual features without supervision.
\newblock \emph{arXiv preprint arXiv:2304.07193}, 2023.

\bibitem[Panachakel \& Ramakrishnan(2021)Panachakel and
  Ramakrishnan]{panachakel2021decoding}
Jerrin~Thomas Panachakel and Angarai~Ganesan Ramakrishnan.
\newblock Decoding covert speech from eeg-a comprehensive review.
\newblock \emph{Frontiers in Neuroscience}, 15:\penalty0 392, 2021.

\bibitem[Papineni et~al.(2002)Papineni, Roukos, Ward, and
  Zhu]{papineni2002bleu}
Kishore Papineni, Salim Roukos, Todd Ward, and Wei-Jing Zhu.
\newblock Bleu: a method for automatic evaluation of machine translation.
\newblock In \emph{Proceedings of the 40th annual meeting of the Association
  for Computational Linguistics}, pp.\  311--318, 2002.

\bibitem[Radford et~al.(2021)Radford, Kim, Hallacy, Ramesh, Goh, Agarwal,
  Sastry, Askell, Mishkin, Clark, et~al.]{radford2021learning}
Alec Radford, Jong~Wook Kim, Chris Hallacy, Aditya Ramesh, Gabriel Goh,
  Sandhini Agarwal, Girish Sastry, Amanda Askell, Pamela Mishkin, Jack Clark,
  et~al.
\newblock Learning transferable visual models from natural language
  supervision.
\newblock In \emph{International conference on machine learning}, pp.\
  8748--8763. PMLR, 2021.

\bibitem[Raffel et~al.(2020)Raffel, Shazeer, Roberts, Lee, Narang, Matena,
  Zhou, Li, and Liu]{raffel2020exploring}
Colin Raffel, Noam Shazeer, Adam Roberts, Katherine Lee, Sharan Narang, Michael
  Matena, Yanqi Zhou, Wei Li, and Peter~J Liu.
\newblock Exploring the limits of transfer learning with a unified text-to-text
  transformer.
\newblock \emph{The Journal of Machine Learning Research}, 21\penalty0
  (1):\penalty0 5485--5551, 2020.

\bibitem[Razavi et~al.(2019)Razavi, Van~den Oord, and
  Vinyals]{razavi2019generating}
Ali Razavi, Aaron Van~den Oord, and Oriol Vinyals.
\newblock Generating diverse high-fidelity images with vq-vae-2.
\newblock \emph{Advances in neural information processing systems}, 32, 2019.

\bibitem[Rombach et~al.(2022)Rombach, Blattmann, Lorenz, Esser, and
  Ommer]{rombach2022high}
Robin Rombach, Andreas Blattmann, Dominik Lorenz, Patrick Esser, and Bj{\"o}rn
  Ommer.
\newblock High-resolution image synthesis with latent diffusion models.
\newblock In \emph{Proceedings of the IEEE/CVF conference on computer vision
  and pattern recognition}, pp.\  10684--10695, 2022.

\bibitem[Singh et~al.(2023)Singh, Pandey, Miyapuram, and
  Raman]{singh2023eeg2image}
Prajwal Singh, Pankaj Pandey, Krishna Miyapuram, and Shanmuganathan Raman.
\newblock Eeg2image: Image reconstruction from eeg brain signals.
\newblock In \emph{ICASSP 2023-2023 IEEE International Conference on Acoustics,
  Speech and Signal Processing (ICASSP)}, pp.\  1--5. IEEE, 2023.

\bibitem[Socher et~al.(2013)Socher, Perelygin, Wu, Chuang, Manning, Ng, and
  Potts]{socher2013recursive}
Richard Socher, Alex Perelygin, Jean Wu, Jason Chuang, Christopher~D Manning,
  Andrew~Y Ng, and Christopher Potts.
\newblock Recursive deep models for semantic compositionality over a sentiment
  treebank.
\newblock In \emph{Proceedings of the 2013 conference on empirical methods in
  natural language processing}, pp.\  1631--1642, 2013.

\bibitem[Sun et~al.(2019)Sun, Wang, Zhang, and Zong]{sun2019towards}
Jingyuan Sun, Shaonan Wang, Jiajun Zhang, and Chengqing Zong.
\newblock Towards sentence-level brain decoding with distributed
  representations.
\newblock In \emph{Proceedings of the AAAI Conference on Artificial
  Intelligence}, volume~33, pp.\  7047--7054, 2019.

\bibitem[Touvron et~al.(2023)Touvron, Martin, Stone, Albert, Almahairi, Babaei,
  Bashlykov, Batra, Bhargava, Bhosale, et~al.]{touvron2023llama}
Hugo Touvron, Louis Martin, Kevin Stone, Peter Albert, Amjad Almahairi, Yasmine
  Babaei, Nikolay Bashlykov, Soumya Batra, Prajjwal Bhargava, Shruti Bhosale,
  et~al.
\newblock Llama 2: Open foundation and fine-tuned chat models.
\newblock \emph{arXiv preprint arXiv:2307.09288}, 2023.

\bibitem[Van Den~Oord et~al.(2017)Van Den~Oord, Vinyals, et~al.]{van2017neural}
Aaron Van Den~Oord, Oriol Vinyals, et~al.
\newblock Neural discrete representation learning.
\newblock \emph{Advances in neural information processing systems}, 30, 2017.

\bibitem[Wang et~al.(2023)Wang, Bao, Dong, Bjorck, Peng, Liu, Aggarwal,
  Mohammed, Singhal, Som, et~al.]{wang2023image}
Wenhui Wang, Hangbo Bao, Li~Dong, Johan Bjorck, Zhiliang Peng, Qiang Liu, Kriti
  Aggarwal, Owais~Khan Mohammed, Saksham Singhal, Subhojit Som, et~al.
\newblock Image as a foreign language: Beit pretraining for vision and
  vision-language tasks.
\newblock In \emph{Proceedings of the IEEE/CVF Conference on Computer Vision
  and Pattern Recognition}, pp.\  19175--19186, 2023.

\bibitem[Wang \& Ji(2022)Wang and Ji]{wang2022open}
Zhenhailong Wang and Heng Ji.
\newblock Open vocabulary electroencephalography-to-text decoding and zero-shot
  sentiment classification.
\newblock In \emph{Proceedings of the AAAI Conference on Artificial
  Intelligence}, volume~36, pp.\  5350--5358, 2022.

\bibitem[Zhang et~al.(2020)Zhang, Zhao, Saleh, and Liu]{zhang2020pegasus}
Jingqing Zhang, Yao Zhao, Mohammad Saleh, and Peter Liu.
\newblock Pegasus: Pre-training with extracted gap-sentences for abstractive
  summarization.
\newblock In \emph{International Conference on Machine Learning}, pp.\
  11328--11339. PMLR, 2020.

\bibitem[Zhou et~al.(2023{\natexlab{a}})Zhou, Duan, Chang, Wang, and
  Lin]{zhou2023beltbootstrapping}
Jinzhao Zhou, Yiqun Duan, Yu-Cheng Chang, Yu-Kai Wang, and Chin-Teng Lin.
\newblock Belt:bootstrapping electroencephalography-to-language decoding and
  zero-shot sentiment classification by natural language supervision.
\newblock \emph{arXiv preprint arXiv:2309.12056}, 2023{\natexlab{a}}.

\bibitem[Zhou et~al.(2023{\natexlab{b}})Zhou, Duan, Zou, Chang, Wang, and
  Lin]{zhou2023speech2eeg}
Jinzhao Zhou, Yiqun Duan, Yingying Zou, Yu-Cheng Chang, Yu-Kai Wang, and
  Chin-Teng Lin.
\newblock Speech2eeg: Leveraging pretrained speech model for eeg signal
  recognition.
\newblock \emph{IEEE Transactions on Neural Systems and Rehabilitation
  Engineering}, 2023{\natexlab{b}}.

\end{thebibliography}
\bibliographystyle{iclr2024_conference}

\newpage

\appendix
\section*{\Large \centering Supplementary Material for BELT-2:\\ 
\emph{Bootstrapping EEG-to-Language representation alignment for multi-task brain decoding}}
 \vspace{20pt}

\section{Related Works}\label{appendix:related-work}
\textbf{EEG decoding} \quad
Prior brain studies demonstrated the potential to decode speech~\citep{anumanchipalli2019speech} and language signals~\citep{anumanchipalli2019speech} from the human brain using invasive neuro-sensors, but the risks make it impractical for most people. More recently, a surge of efforts was made to extract rich information from noninvasive brain signals through advanced representation learning techniques, opening the door to a wide array of innovative tasks based on brain signals, such as image reconstruction~\citep{singh2023eeg2image} and movement prediction~\citep{zhou2023speech2eeg}. Nonetheless, Many of these efforts have limitations, including vocabulary size and decoding performance, hindering their suitability for complex practical scenarios.
Our work focuses on open-vocabulary sentence decoding from noninvasive brain signals with fluent decoding performance and versatile multi-task adaptability, making it a promising solution for a diverse range of applications.

\textbf{EEG-Language representation alignment} \quad
A crucial step for most cross-modality tasks is the acquisition of aligned multi-modal representations \citep{liu2023audioldm, mokady2021clipcap, rombach2022high}. Achieving this involves an alignment step following the acquisition of unimodality pretrained models \citep{li2023blip}. Yet, the formidable challenge persists due to the limited scale and sparse EEG dataset annotations, as we strive to create a semantically coherent and universally adaptable EEG encoder, akin to visual counterparts \citep{dosovitskiy2020image,radford2021learning}.

Diverging from the conventional fully-supervised paradigm, infusing natural language supervision enriches non-language modalities representation with semantics and zero-shot generalization~\citep{desai2021virtex}. Previous studies in unimodal vision tasks show that a large vision encoder, trained directly with language supervision, can match performance compared to learning from massive datasets \citep{joulin2016learning}. Recent works incorporating language-guided learning also support the value of additional semantics for non-language representation generalization~\citep{wang2023image,elizalde2023clap}. Inspired by their successes, our work endeavors to bootstrap the learning of an Encoder that aligns EEG and language representation through natural language supervision.

\section{Mathematical symbols used in this paper}
In Table \ref{appendix:math} we show a list of mathematical symbols used in this paper.
\begin{table}[h!]
\centering
\caption{List of mathematical symbols used in this paper\label{appendix:math}}
\begin{tabular}{c|l|c|l}
\hline
Symbol                                          & \multicolumn{1}{c|}{Description} & Symbol           & \multicolumn{1}{c}{Description}     \\ \hline
\multirow{2}{*}{$\left\langle\mathcal{E},\mathcal{S}\right\rangle$} & Word-level EEG embedding & \multirow{2}{*}{$\left\langle\mathcal{E},c\right\rangle$} & Word-level EEG embedding \\
                                                & sequence and text sentence pair  &        & sequence and sentiment label pair   \\ \hline
\multirow{2}{*}{$\left\langle\mathcal{E},\hat{\mathcal{S}}\right\rangle$} & Word-level EEG embedding& $\mathbf{w}\in\mathcal{W}$   & BPE text token's embeddings \\\cline{3-4} 
                                                & sequence and text summary pair   & $\mathbf{e}\in\mathcal{E}$    & EEG embedding vector    \\ \hline
$c\in\mathcal{C}$                               & Sentiment label                  & $\mathbf{v}\in\mathcal{V}$    & Discrete codebook embeddings    \\ 
\hline
\end{tabular}
\end{table}

\section{Implementation Details}\label{appendix:implementation}
\subsection{Implementation Details for the Q-Conformer}

The Q-Conformer is implemented using the configuration detailed in Table \ref{tab:conformer-structure}. The detailed structures for the convolution module are shown in Table \ref{tab:convolution-module-structure}. We use the same Conformer block for the encoder and decoder, each with $2$ Conformer blocks. We trained All models are trained on Nvidia A40 GPUs. 
\begin{table}[]
\caption{Detailed configuration of the conformer block \label{tab:conformer-structure}}
\centering
\begin{tabular}{llll}
\hline
Layer                                                        & Hidden Size & Activation Function & Number of Heads \\\hline
Layer Norm                                                   & 840         & -                   & -              \\
Feed Forward Module                                          & 840         & GELU                & -              \\
LayerNorm                                                    & 840         & -                   & -              \\
Multi-Head Self Attention                                    & 840         & -                   & 8              \\
\begin{tabular}[c]{@{}l@{}}Convolution\\ Module\end{tabular} & 840         & -                   & -              \\
Layer Norm                                                   & 840         & -                   & -              \\
Feed Forward Module                                          & 840         & GELU                & -              \\
LayerNorm                                                    & 840         & -                   & -             \\\hline
\end{tabular}
\end{table}

\begin{table}[]
\caption{Detailed configuration of the convolution module  \label{tab:convolution-module-structure}}
\centering
\begin{tabular}{l|cccc}
\hline
Layer                 & \multicolumn{1}{l}{Kerrnel} & \multicolumn{1}{l}{Stride} & \multicolumn{1}{l}{In Channel} & \multicolumn{1}{l}{Out Channel} \\ \hline
Layer Norm            & -                           & -                          & 840                            & 840                             \\
Pointwise Convolution & 1                           & 1                          & 840                            & $2\times{840}$                    \\
Depthwise Convolution & 31                          & 1                          & 840                            & 840                             \\
Batch Norm            & -                           & -                          & 840                            & 840                             \\
Pointwise Convolution & 1                           & 1                          & 840                            & 840                             \\
Dropout               & -                           & -                          & -                              & -                               \\ \hline
\end{tabular} 
\end{table}

\subsection{Training details for EEG-to-language alignment learning}
To train the Q-Conformer during the EEG-to-language alignment learning, we use a weighted summation of all the following loss terms:
\begin{equation}\label{eq:task-translation-all}
\mathcal{L}=\lambda_1\mathcal{L}_{vq}+ \lambda_2\mathcal{L}_{bpe} +  \lambda_4\mathcal{L}_{elm} + \lambda_3\mathcal{L}_{neg},
\end{equation}
$\lambda_1$ to $\lambda_4$ are coefficients for each loss term. We set $\lambda_1$ to $\lambda_4$ as $[1,10,10,0.001]$. The main reason for such a setting is the aim to prioritize the learning of achieving EEG-to-language alignment and the training of the query prompt specific to the ELM task. To avoid collapse in training, we implemented the gradient normalization method to normalize the scale of the loss function and stabilize the training process. 

\subsection{Training virtual prefix for bridging Q-Conformer and LLM}\label{sec:prefixtuning}
The prefix-tuning method used in our paper closely follows the implementation in \cite{li2021prefix}, the objective function ($\mathcal{L}_{bridge}$) is defined as a modified loss function tailored to guide the selective of continuous virtual prefix prompts. We use $\theta$ to denote the matrix that stores the virtual prefix. Using the machine translation loss $\mathcal{L}_{tr}$ as an example, the objective function can be expressed as:
\begin{equation}
\mathcal{L}(\theta_{\text{bridge}}) = \mathcal{L}_{tr}(\hat{\mathcal{S}}, \mathcal{S})
\end{equation}
In this example, the prefix prompts to learn properly describe the EEG-to-Langugage translation task to the subsequence frozen LLM, utilizing the generation capacity of the LLM models to improve translation performance. 

\subsection{Training details for multi-task learning}
To extend our model to multi-task decoding, we simultaneously train the model in three EEG decoding tasks including translation, summary, and sentiment classification task. We randomly sample a task for each batch during the training epochs. The loss function for translation task $\mathcal{L}_{tr}$ and sentiment classification tasks $\mathcal{L}_{st}$ are illustrated in Equation \ref{eq:tranlation-task-1} and Equation \ref{eq:sst-task} respectively. 

For learning the summary task, the loss function could be written as follows:
\begin{equation}\label{eq:lm-loss-sum}
\mathcal{L}_{sum} =-\sum^{\vert\hat{\mathcal{S}}\vert}_l\log{p(s_l\in\hat{\mathcal{S}})} 
\end{equation}
, where $p(s_l)$ denotes a model predicting the word token for the next location. The final multi-task objective $\mathcal{L}$ is written as follows:
\begin{equation}\label{eq:lm-loss-sum}
\mathcal{L}_{mt} =\mathcal{L}_{tr} + \mathcal{L}_{sum} + \mathcal{L}_{st} 
\end{equation}

\section{Improved Q-Conformer EEG Encoder}
\begin{figure}[h!]
    \centering
    \includegraphics[width=\linewidth]{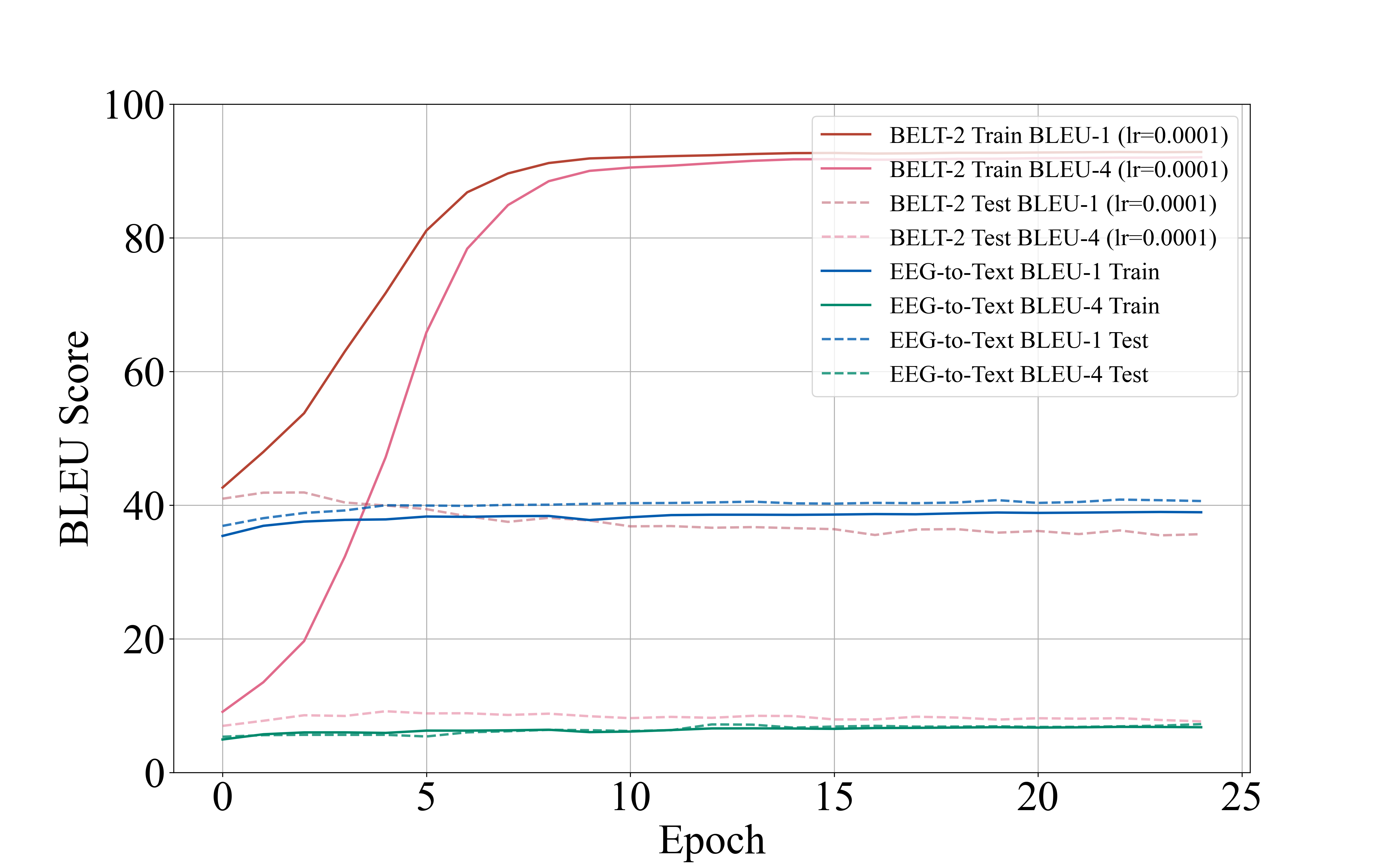}
    \caption{EEG encoder performance comparison}
    \label{fig:EncoderPerformance}
\end{figure}

We observed a noteworthy trend when utilizing a relatively larger learning rate of $1e-4$, as opposed to the optimal learning rate of $5e-6$ for the top-performing Q-Conformer Encoder, as indicated in Figure \ref{fig:EncoderPerformance}. This variance in learning rates led to a remarkable performance by the Q-Conformer Encoder on the training dataset, resulting in notably high BLEU Scores. Specifically, the BLEU-$1$ and BLEU-$4$ scores soared to remarkable levels, reaching $93.03$ and $92.69$ respectively. In stark contrast, the EEG-to-Text baseline method significantly lagged behind, registering only BLEU-${1,4}$ scores of $38.98$ and $6.82$ during our replicated training, highlighting the superior EEG encoding capabilities of the Q-Conformer Encoder.

It's also worth noting that the BLEU-1 performance of the Q-Conformer encoder experienced a decline from $42.43$ to $35.48$ during the testing phase, we interpret this as a minor setback. Such a reduction in performance can often be attributed to the challenges of generalization, which frequently happen in the context of training on a relatively small dataset.

Furthermore, it's worth highlighting that within this setting, the Q-Conformer still achieved a testing BLEU-4 score of $9.3$, surpassing the baseline EEG-to-Text method's training set BLEU-4 score. This outcome serves as a compelling testament to the enhanced encoding capacity conferred by our Q-Conformer Encoder.

\section{Comparison with BELT-2 without BPE-level Contrasive Learning}
\begin{figure}[h!]
    \centering
    \subfigure[]{
        \includegraphics[width=0.45\linewidth]{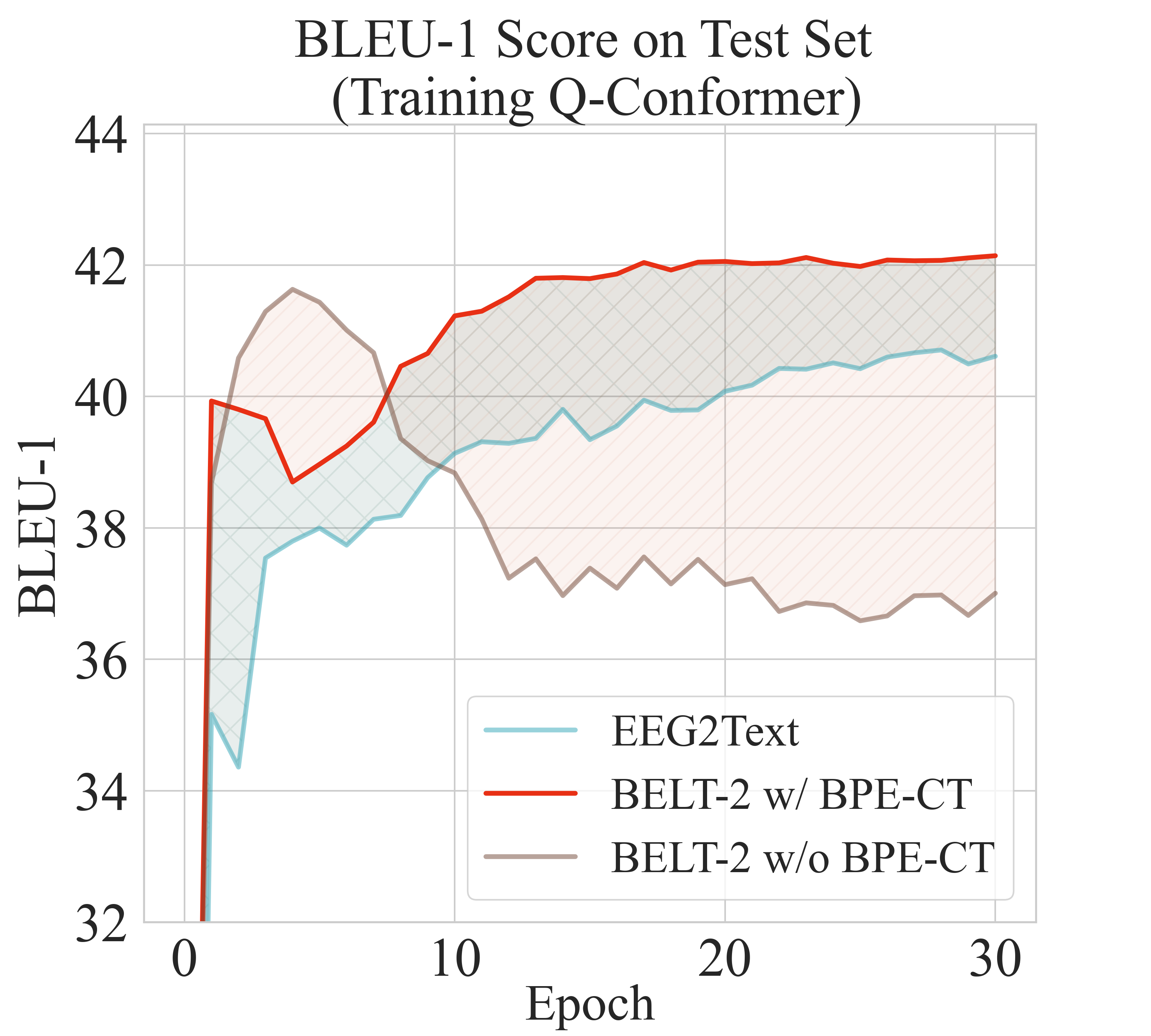}
        \label{fig:compare_bleu1}
    }\hspace{-6mm}\vspace{-1mm}
    \subfigure[]{
        \includegraphics[width=0.45\linewidth]{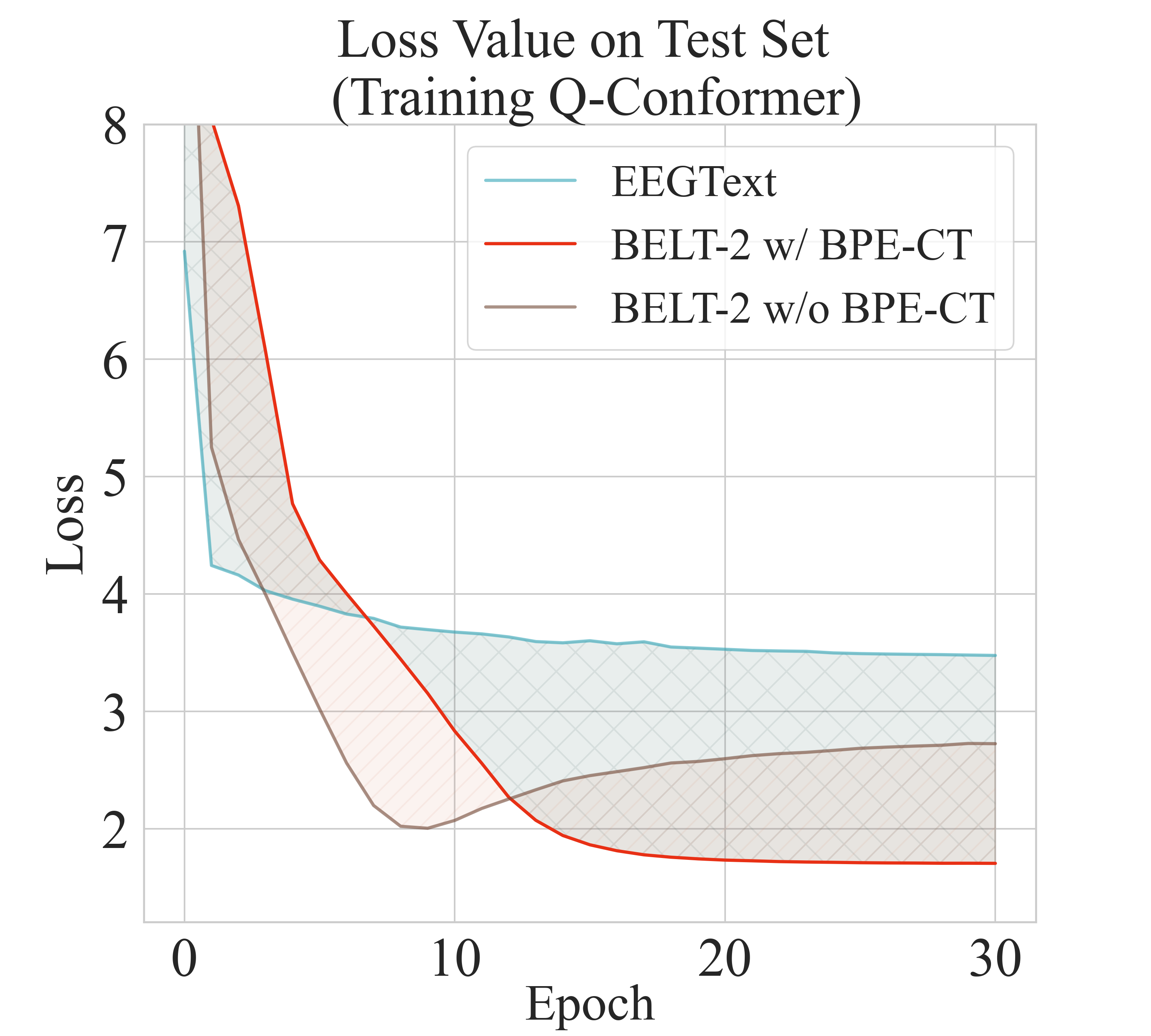}
         \label{fig:compare_loss}
    }\hspace{-6mm}\vspace{-1mm}
    \caption{Ablation Study on Different Settings\label{fig:BLEU-Com}}
\end{figure}

In Figure \ref{fig:compare_bleu1} and Figure \ref{fig:compare_loss}, we present a comprehensive comparison of the learning curves and BLEU-1 curve of the baseline EEG-to-Text model \citep{cruttenden2014gimson}, the Q-Conformer encoder without applying the BPE-level contrastive learning (BELT-2 w/o BPE-CT) and the Q-Conformer encoder with BPE-level contrastive learning (BELT-2 w/ BPE-CT)g. The visualized learning curves include the BLEU-1 score and loss values for $30$ epochs on the test split. Comparing the EEG-to-Text model and the BELT-2 model, it's evident that BELT-2 offers a significant reduction in loss values with or without BPE-level contrastive learning, indicating the proposed model architecture is more efficient in capturing EEG patterns. However, a notable observation arises after epoch $8$. Without the BPE-contrastive learning (orange curves), the BLEU-1 score fluctuates and drops significantly.  On the contrary, the introduction of BPE-level loss helps stabilize the model's performance, particularly on unseen EEG data. This highlights the substantial enhancement brought about by our proposed BPE-contrastive learning framework.

\section{Multi-task training results} \label{Appendix:multitask-training}
\begin{figure}[h!]
    \centering
    \subfigure[Multi-task training \textbf{without} pretrained Q-Comformer Weights]{
        \includegraphics[width=0.46\linewidth]{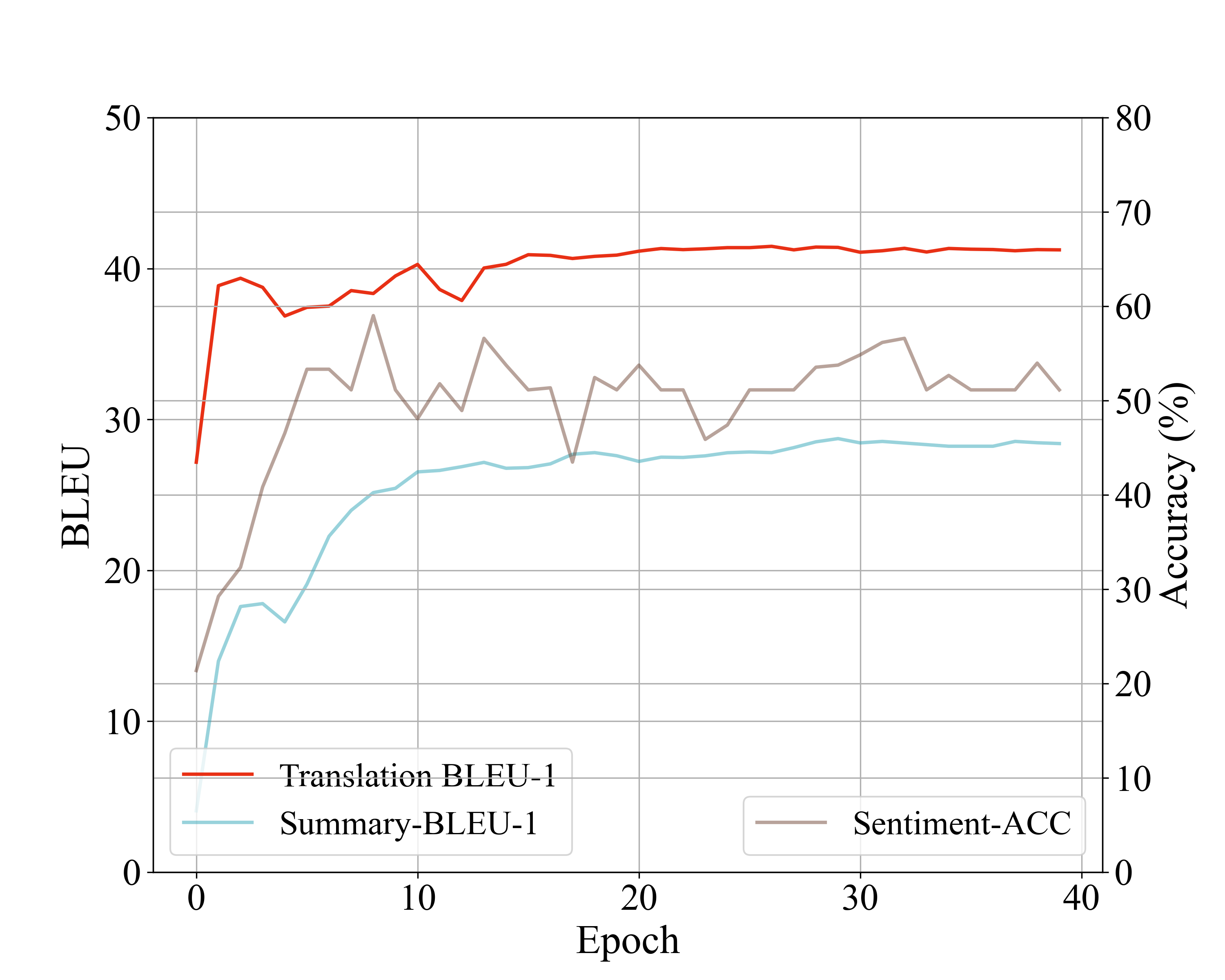}
        \label{fig:multitask-no-pretrained}
    }\hspace{0mm}\vspace{0mm}
    \subfigure[Multi-task training \textbf{with} pretrained Q-Comformer Weights]{
        \includegraphics[width=0.46\linewidth]{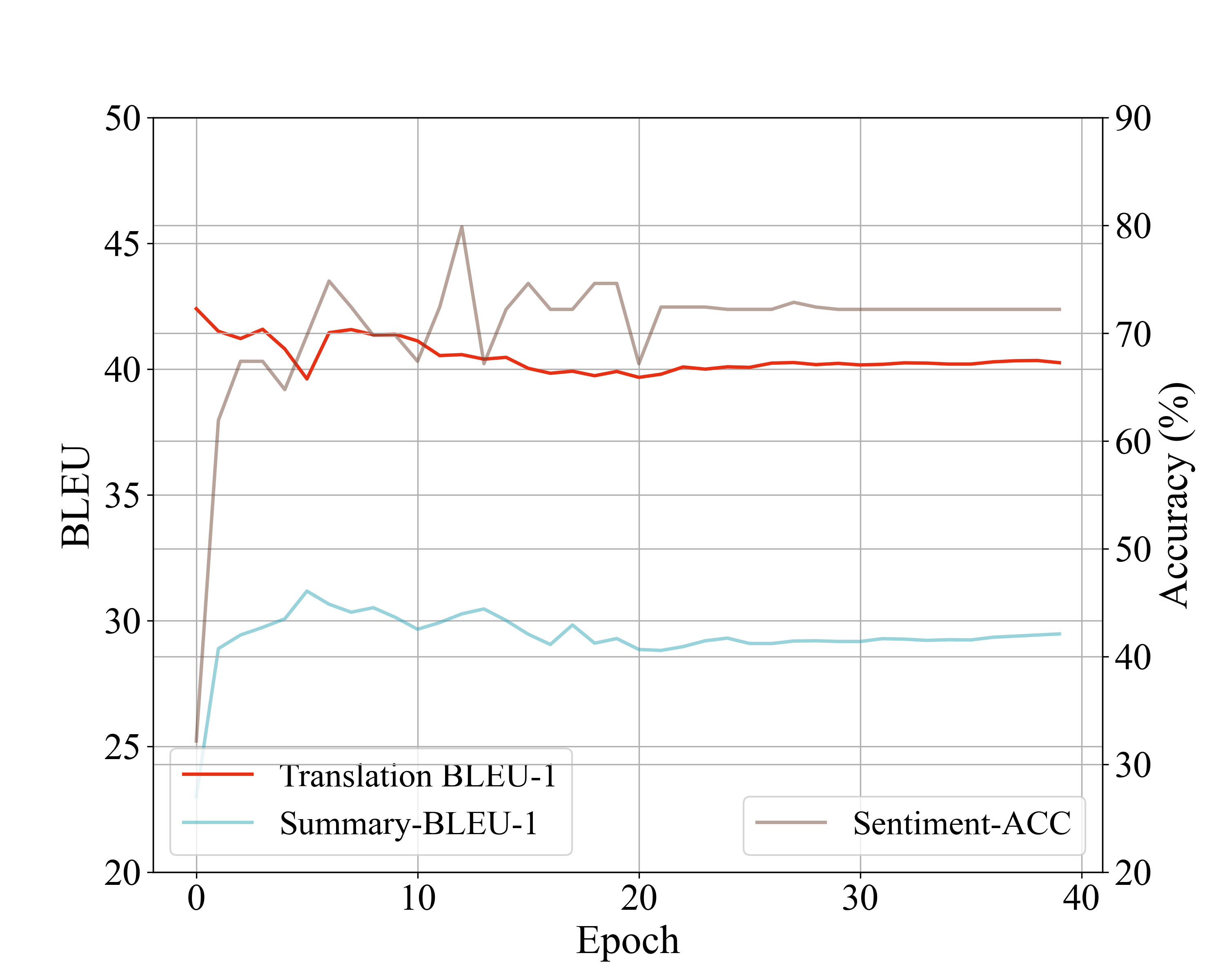}
     \label{fig:multitask-has-pretrained}
    }\hspace{0mm}\vspace{0mm}    
    \caption{Ablation study on multitask learning and effect of our pretrained weights}
    \label{Multitask-metrics}
\end{figure}

We show the performance of translation, summary, and sentiment classification on the test set during the multitask training learning phase of BELT-2 in Table \ref{Multitask-metrics}. In Table \ref{fig:multitask-no-pretrained}, we can observe that without the use of pretrained weights, all tasks are learned from scratch. In this case, the translation BLEU-1 score starts from $4.06$ BLEU-1 score and rises to only reaches $41.47$ and the summarization BLEU-1 score reaches $28.72$. Also, the sentiment classification accuracy gradually increased to $59\%$. However, the use of Q-Conformer pretrained on translation tasks could improve the training stability and performance of both the sentiment classification task and the summarization task. Due to the pretrained weights, we observed that in Table \ref{fig:multitask-has-pretrained}, the BLEU-1 score of the summarization performance and sentiment achieved $23.0$ BLEU-1 score after the first training epoch. Then continued to increase to $31.17$. The accuracy for sentiment classification also reaches $79.86\%$ at its peak and stabilizes at around $74\%$. However, the performance of the translation task slightly decreased. This is an expected phenomenon in multi-task training. Nonetheless, this ethernet still shows the multi-task learning capacity and extensibility of our BELT-2 framework. 

\section{Generated Summarization Results}
We created the summarization dataset with the prompt "Rewrite the sentence by summarizing its main idea using $8$ words from the sentence and keep the summarized sentence similar to the original sentence: $\{s\}$" where and $\{s\}$ is the original sentence from the dataset. Table \ref{tab:summary-all-set} showcases summary and prediction samples generated by the BELT-2 model. We could see those summary ground truths cover the key ideas of the original sentence and are within the maximum summarization word limit.  On the training set, our BELT-2 model could learn and precisely generate a summary of the EEG signal, such as "film with twists" vs. "film with twists.". However, this summarization capacity did not generalize well on unseen test and validation data. We consider the lack of training data as one of the major reasons for this problem. Another reason is that our current model lacks higher-level skill that requires additional reasoning and abstraction skills beyond the mere translation of the brain signal, which leaves room for future improvements. 
\begin{table}[h!]
\caption{Summarization examples and generated results on the train set. The \textbf{bold} denotes an exact match between the ground truth and our prediction. \uline{underline} denotes a fuzzy match with similar semantic meanings. \label{tab:summary-all-set} }
\begin{tabular}{lll}
\hline
\multicolumn{1}{c}{\textbf{}} & \multicolumn{2}{c}{\textbf{Training}} \\ \hline
(1)                           & \multirow{3}{*}{Sentence} & Beautifully crafted, engaging filmmaking that should attract                 \\
                              &                           & upscale audiences hungry for quality and a nostalgic, twisty yarn            \\
                              &                           & that will keep them guessing.                                                \\
                              & Summary GT                & High\textbf{-quality film with twists.}                                               \\
                              & Prediction                & \textbf{-quality film with twists.}                                                   \\ \hline
(2)                           & Sentence                  & Slow, silly and unintentionally hilarious.                                   \\
                              & Summary GT                & Silly, \uline{\textbf{slow} comedy}.                                                          \\
                              & Prediction                & inger, \uline{\textbf{slow} movie}.                                                           \\ \hline
(3)                           & \multirow{2}{*}{Sentence} & The movie is for fans who can't stop loving anime, and the                   \\
                              &                           & fanatical excess built into it.                                              \\
                              & Summary GT                & Anime \textbf{fans will love} excessive movie.                                        \\
                              & Prediction                & imated \textbf{fans will love}  this gore.                                           \\ \hline
(4)                           & Sentence                  & But here's the real damn: It isn't funny, either.                            \\
                              & Summary GT                & Funny, but \textbf{not really.}                                                       \\
                              & Prediction                & unny, smart \textbf{not really. }                                                     \\ \hline
(5)                           & \multirow{2}{*}{Sentence} & Everything was as superficial as the forced New Jersey                       \\
                              &                           & lowbrow accent Uma had.                                                      \\
                              & Summary GT                & Uma's \textbf{accent was fake.}                                                       \\
                              & Prediction                & ma's \textbf{accent was fake.}                                                  \\ \hline
(6)                           & \multirow{2}{*}{Sentence} & Feels like nothing quite so much as a middle-aged moviemaker's               \\
                              &                           & attempt to surround himself with beautiful, half-naked women.                \\
                              & Summary GT                & Filmmaker surrounds \textbf{himself with beautiful women.}                            \\
                              & Prediction                & mmakers imagined   \textbf{himself with beautiful women.}                             \\ \hline
(7)                           & Sentence                  & He died in Springport, New York in 1815.                                     \\
                              & Summary GT                & Man \textbf{passed away in Springport. }                                              \\
                              & Prediction                & \textbf{passed away in   Springport. }                                                \\ \hline
\multicolumn{1}{c}{}          & \multicolumn{2}{c}{\textbf{Test and Validataion}}                                                        \\ \hline
(1)                           & \multirow{2}{*}{Sentence} & A richly imagined and admirably mature work from a gifted                    \\
                              &                           & director who definitely has something on his mind.                           \\
                              & Summary GT                & Director's mature work reflects deep thoughts.                               \\
                              & Prediction                & 's debut film. his   empathy.                                                \\ \hline
(2)                           & \multirow{2}{*}{Sentence} & An amateurish, quasi-improvised acting exercise shot on                      \\
                              &                           & ugly digital video.                                                          \\
                              & Summary GT                & \textbf{Ugly} video showcases poor \textbf{acting}.                                            \\
                              & Prediction                & ma,, \textbf{ugly} \textbf{acting}.                                                            \\ \hline
(3)                           & \multirow{3}{*}{Sentence} & Warm Water Under a Red Bridge is a quirky and poignant                       \\
                              &                           & Japanese film that explores the fascinating connections                      \\
                              &                           & between women, water, nature, and sexuality.                                 \\
                              & Summary GT                & Japanese film explores women, water, nature, sexuality   poignantly.         \\
                              & Prediction                & actor, themes's love,   and. love.eticsancy.                                 \\ \hline
(4)                           & Sentence                  & It just doesn't have much else... especially in a moral sense.               \\
                              & Summary GT                & Limited moral \textbf{compass}                                                        \\
                              & Prediction                & role \textbf{compass}.                                                                \\ \hline
(5)                           & Sentence                  & It's solid and affecting and exactly as thought-provoking as   it should be. \\
                              & Summary GT                & Thought-\textbf{provoking} and solid.                                                 \\
                              & Prediction                & inful\textbf{provoking} film   funny.                                                 \\ \hline
(6)                          & \multirow{4}{*}{Sentence} & The art direction is often exquisite, and the anthropomorphic animal         \\
                              &                           & characters are beautifully realized through clever makeup design,            \\
                              &                           & leaving one to hope that the eventual DVD release will                       \\
                              &                           & offer subtitles and the original Italian-language soundtrack.                \\
                              & Summary GT                & Beautiful animal characters, DVD subtitles.                                  \\
                              & Prediction                & iful, inter. funny   experience.                                             \\ \hline

\end{tabular}
\end{table}

\section{Ablation Experiments on Hyper-Parameters}\label{Appendix-hyperparameter}
\begin{figure}[h!]
    \centering
    \subfigure[Learning Rate]{
        \includegraphics[width=0.45\linewidth]{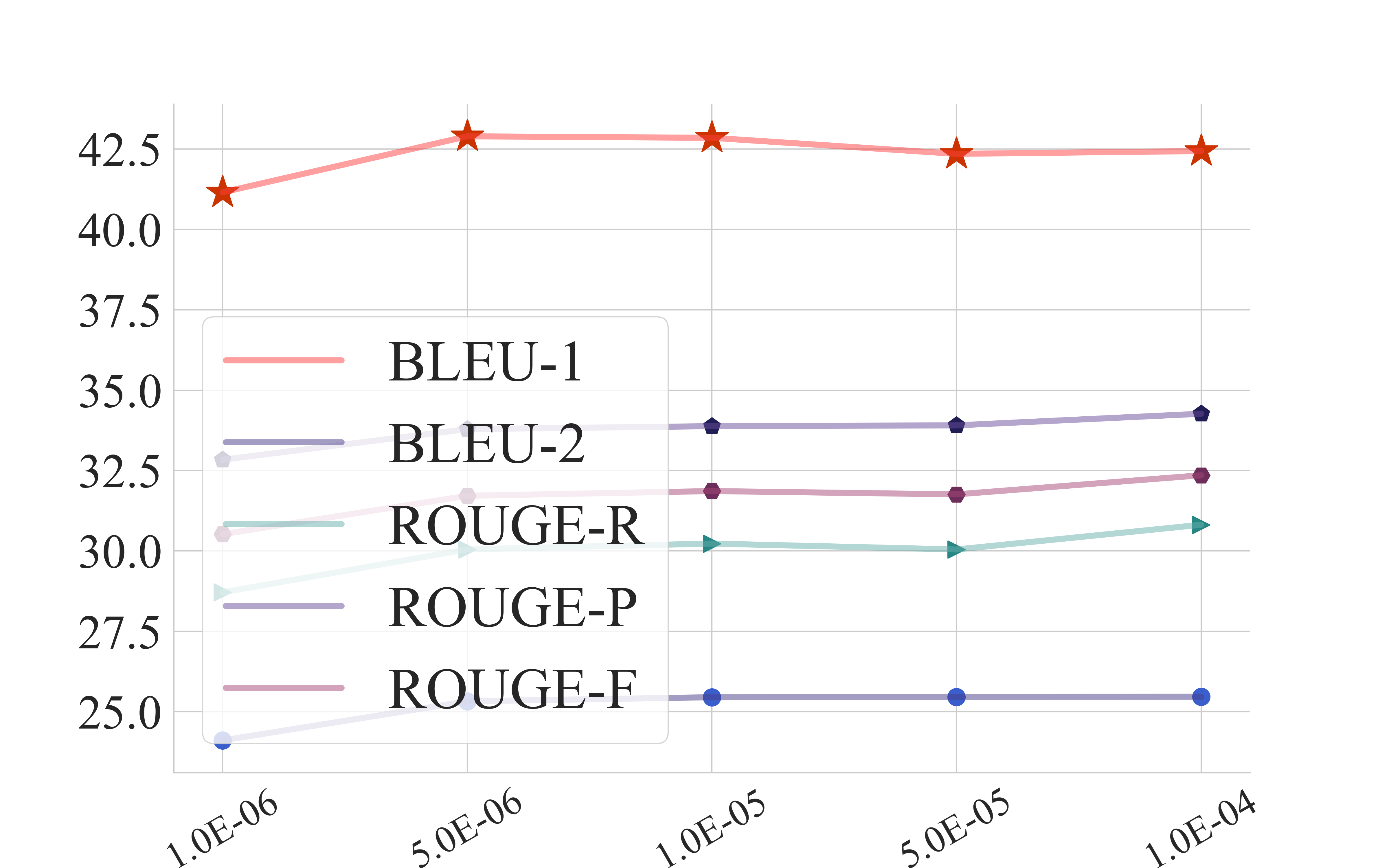}
        \label{fig:ablation-lr}
    }\hspace{0mm}\vspace{0mm}
    \subfigure[Batch Size]{
    	\includegraphics[width=0.45\linewidth]{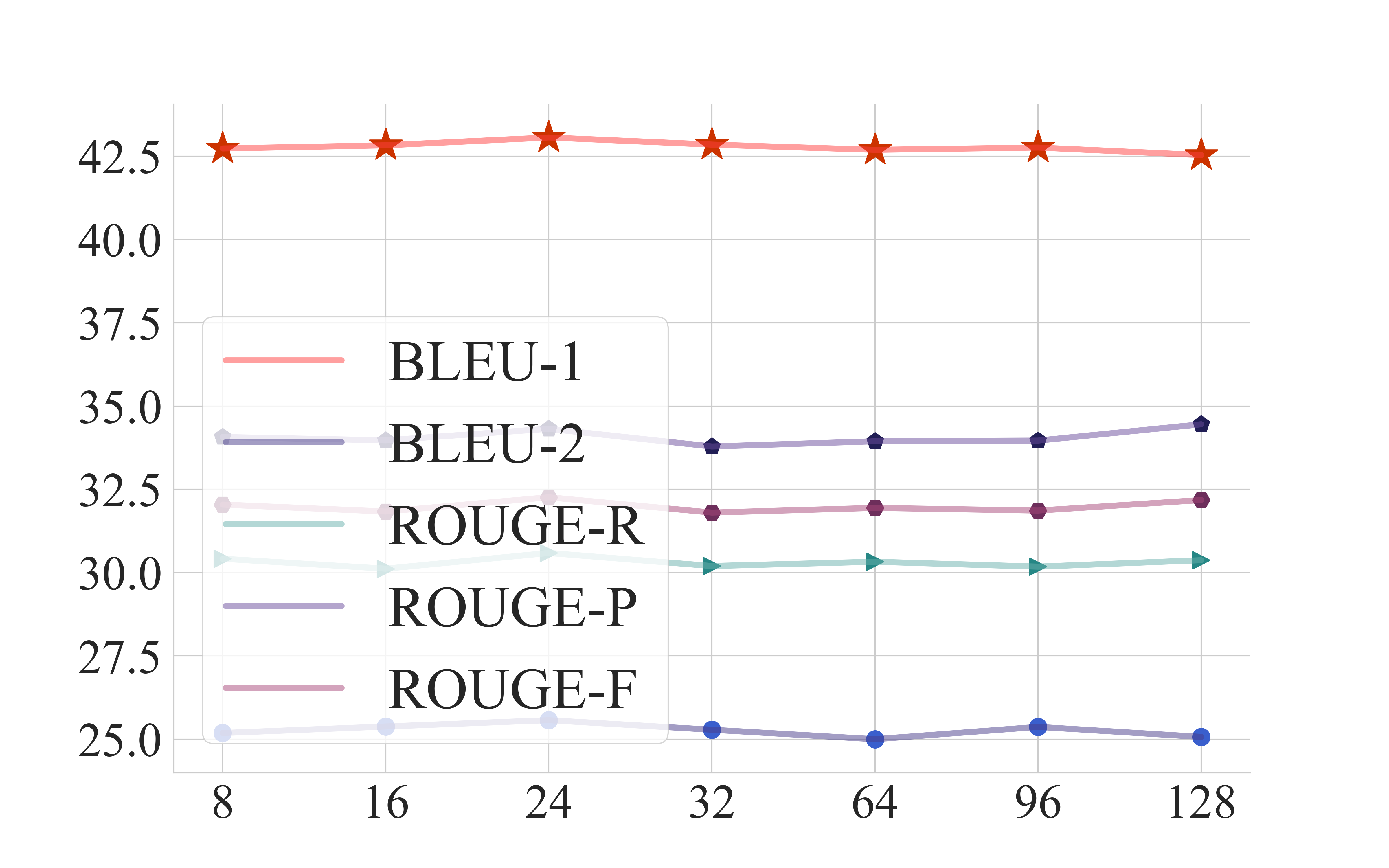}
     \label{fig:ablation-bs}
    }\hspace{0mm}\vspace{0mm}
    \subfigure[Cross-attention frequency]{
    	\includegraphics[width=0.45\linewidth]{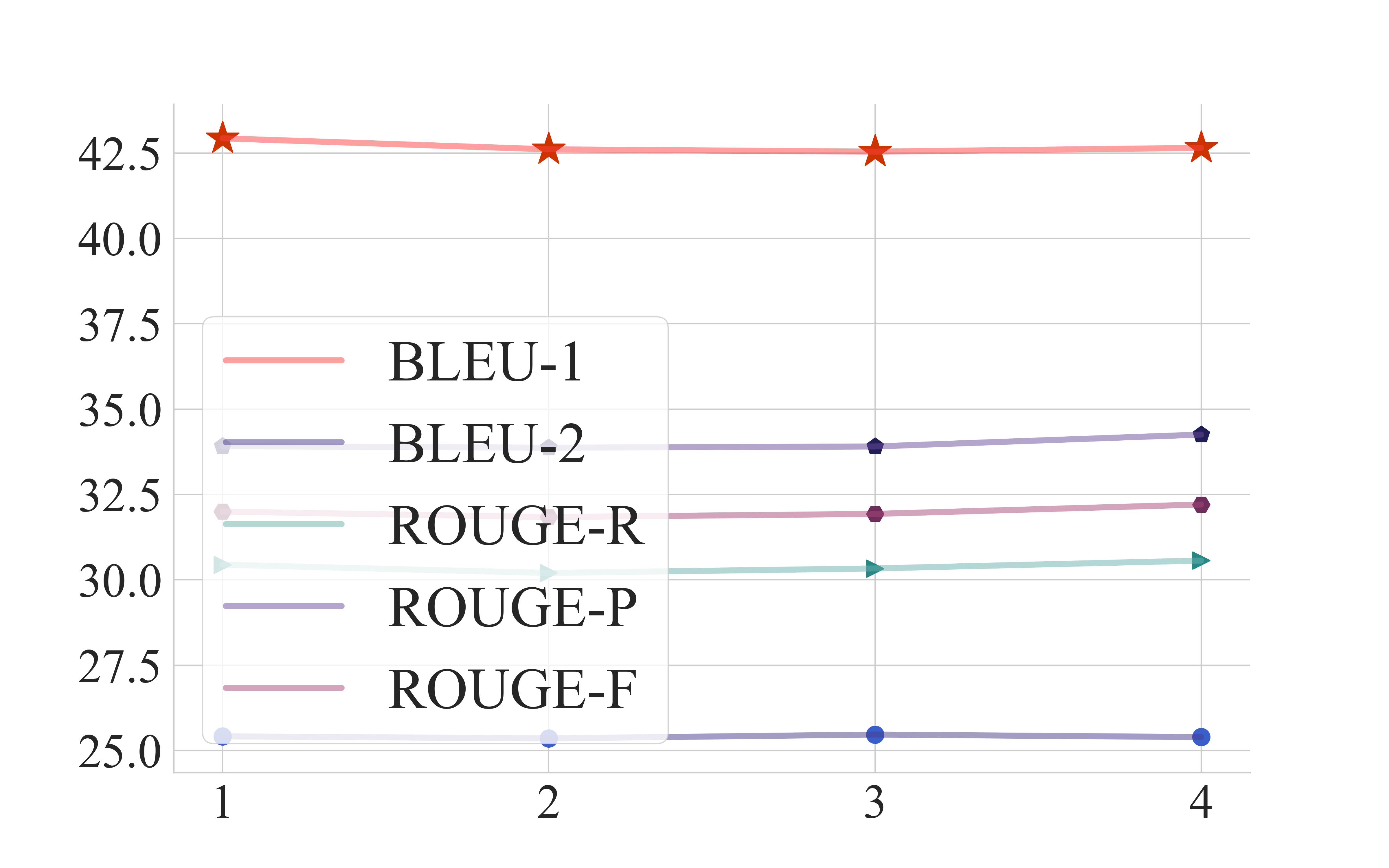}
     \label{fig:ablation-freq}
    }\hspace{0mm}\vspace{0mm}
    \subfigure[Query Prompt Size]{
	\includegraphics[width=0.45\linewidth]{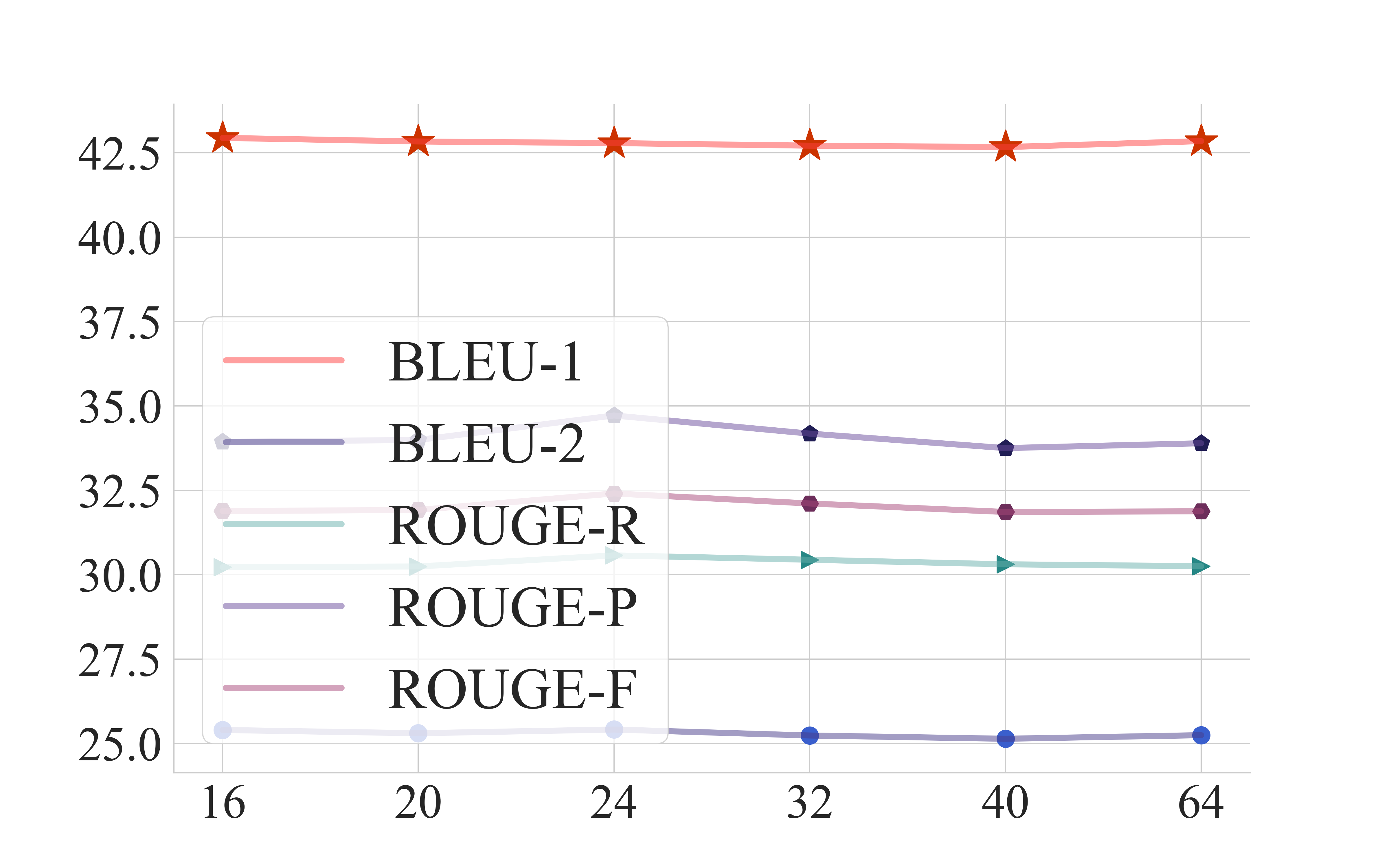}
     \label{fig:ablation-token}
    }\hspace{0mm}\vspace{0mm}
    \caption{Ablation study on hyper-parameters.}
    \label{Ablation-metrics}
\end{figure}

We conducted an ablation study on different hyper-parameters including the learning rate, batch size, frequency of the inserted cross-attention layer in the context layer of the Q-Conformer, and the number of querying prompts. The evaluation metrics can be found in Figure \ref{Ablation-metrics}. We observe that the introduction of BPE-contrastive learning consistently improves training stability and model performance in different hyper-parameter settings. This result shows that the learning performance of BELT-2's EEG encoder is not easily affected by the change of training parameters and is relatively easy to reproduce. 

\section{Augmentation effect of Speculative Augmentation}\label{Appendix-sa}
The limitation of unique sentence from the training dataset also limits the diversity of the MLC context outputed by the Q-Conformer. The training set we used in our cross-sentence setting contains only $790$ unique sentences as target for prefix-tuning when bridging Q-Conformer and LLM. For the Q-Conformer, predicts around $900$ uniques MLC throughout the training dataset. This lack of training inputs makes the training for a good virtual prefix difficult. To solve this problem, our speculative augmentation method reuse cached MLC from the training stage of Q-Coformer. When using MLC from $K=15$ checkpoints, we achieve a total of $5107$ samples for prefix-tuning.  

\section{Extensive examples of generated translation outputs}\label{Appendix-moretrans}
We provide extensive translation outputs from our BELT-2 model compared with the baseline EEG-to-Text model and the ground truth in Table \ref{many-results}. It shows that for some samples, the BELT-2 model still has insufficient performance, which indicates room for future improvements. 

\begin{table}[]
\caption{Extensive examples of generated translation outputs from unseen EEG signals in the test set. The \textbf{bold} denotes an exact match while \uline{underline} denotes a fuzzy match with similar semantic meanings.\label{many-results}}
\begin{tabular}{lll}
\hline
(1) & Target                  & \textbf{It's not a} particularly good film, \textbf{but} neither \textbf{is it} a monsterous \textbf{one}.    \\ \cline{2-3} 
    & Others                  & was a a bad good story, but it is it bad bad. one.                                                            \\ \cline{2-3} 
    & Ours                    & \underline{\textbf{It's not a} bad bad movie}, \underline{but it \textbf{is it} kinda good bad \textbf{one}}. \\ \hline
(2) & Target                  & It's solid and affecting and exactly as thought-\textbf{provoking as} it should \textbf{be}.                  \\ \cline{2-3} 
    & Others                  & was a, it, it what it.provoking as it is be.                                                                  \\ \cline{2-3} 
    & Ours                    & \underline{\textbf{It's}, believable}, is what -\textbf{provoking as} the sounds \textbf{be}.                 \\ \hline
(3) & \multirow{2}{*}{Target} & Co-writer/director Jonathan Parker's attempts to fashion a Brazil-like,                                       \\
    &                         & hyper-real satire fall dreadfully short.                                                                      \\ \cline{2-3} 
    & \multirow{2}{*}{Others} & operfounder of\textbf{director} of Dem is novel to make a film-themed film                                             \\
    &                         & but-realistic of flatfully short of                                                                           \\ \cline{2-3} 
    & \multirow{2}{*}{Ours}   & Theenstarrings\textbf{director} John Dem hass films to make a new-style                                                \\
    &                         & film -realisticromre are flatareadfully flat.                                                                 \\ \hline
(4) & \multirow{3}{*}{Target} & \textbf{After} World\textbf{ War II,} Kennedy \uline{entered politics} (\uline{partly to fill \textbf{the void}} of his                                  \\
    &                         & popular brother, Joseph P. \textbf{Kennedy}, Jr., on whom his family                                                   \\
    &                         & had pinned many \textbf{of their hopes} but who was killed \uline{in the war}).                                                \\ \cline{2-3} 
    & \multirow{3}{*}{Others} & the War II, the was the andasly as serve the void left a father father                                        \\
    &                         & , John Kennedy. \textbf{Kennedy}, who.) who the he father had been                                                     \\
    &                         & their of his \textbf{hopes}). never was never in the war).                                                             \\ \cline{2-3} 
    & \multirow{3}{*}{Ours}   & \textbf{After the War II,} \uline{became politics},andly \uline{to fulfill\textbf{ the void}}                                                   \\
    &                         & left his father father, John \textbf{Kennedy}. Kennedy, who.,who the Kennedy                                           \\
    &                         & family had placedbased their \textbf{of their hopes}). had had \uline{in Battle}.                                              \\ \hline
(5) & Target                  & \textbf{It's} solid and affecting and exactly as thought-\textbf{provoking} as it should be.                                    \\ \cline{2-3} 
    & Others                  & was a, it, it what it.outoking as the sounds be.                                                              \\ \cline{2-3} 
    & Ours                    & \textbf{It's}, logical, is what -\textbf{provoking} as the sounds be.                                                           \\ \hline
(6) & \multirow{2}{*}{Target} & \textbf{Too much} of this well-acted but dangerously slow thriller feels like a preamble                               \\
    &                         & to a bigger, more complicated story, one that never materializes.                                             \\ \cline{2-3} 
    & \multirow{2}{*}{Others} & bad of a is-known, not over- is like a film-ble to a more, more dramatic story.                               \\
    &                         & which that will quiteizes.                                                                                    \\ \cline{2-3} 
    & \multirow{2}{*}{Ours}   & \textbf{Too much} drama is-made, unly un-. like a -ble to a much, more serious,.                                       \\
    &                         & one that' quiteizes.                                                                                          \\ \hline
(7)  & Target                & \uline{In 1923} \textbf{he was awarded the} inaugural Bôcher Memorial\textbf{Prize} by               \\
     &                       & the American Mathematical \textbf{Society}.                                          \\ \cline{2-3} 
     & Others                & the, married born the Nobel Pulitzericentne \textbf{Prize} Medal for the             \\
     &                       & French Academyical \textbf{Society}.                                                 \\ \cline{2-3} 
     & Ours                  & \uline{In 1815},\textbf{he was awarded the} Pulécher Prize \textbf{Prize}, the                        \\
     &                       & Royal Academyematical \textbf{Society}.                                              \\ \hline
(8) & Target                & \textbf{He later became} an educator, teaching music theory \textbf{at the University of}     \\
     &                       & the District of Columbia; he was also director of the District of           \\
     &                       & \textbf{Columbia} Music Center jazz workshop band.                                   \\ \cline{2-3} 
     & Others                & was \textbf{became} a actor and and at and and the University of                     \\
     &                       & California Arts of Columbia. and also also a of the                         \\
     &                       & University of \textbf{Columbia}'s School. department..                               \\ \cline{2-3} 
     & \multirow{2}{*}{Ours} & \textbf{He later became} associate at and at at \textbf{at the University of}                 \\
     &                       & California West of Columbia and and he also of the                           \\
     &                       & English' \textbf{Columbia}' Department. department.                                  \\ \hline
(9) & Target                & \textbf{Fans of the} \uline{TV series} \textbf{will be} disappointed, and everyone                    \\
     &                       & else \uline{will be slightly bored.}     \\ \cline{2-3} 
     & Others                & of the film show " remember familiar to however the will                    \\
     &                       & will be happy amused.                                                       \\ \cline{2-3} 
     & Ours                  & \textbf{Fans of the} \uline{movie series} \textbf{will be}, as the who \uline{will}                           \\
     &                       & \uline{be left disappointed.}                                                       \\ \hline
\end{tabular}
\end{table}


\end{document}